\documentclass[12pt]{article}
\usepackage{graphicx}
\def\hybrid{\topmargin 0pt      \oddsidemargin 0pt
        \headheight 0pt \headsep 0pt
       \voffset-1cm
        \textwidth 6.25in       % A4 paper
       \textheight 9.5in       % A4 paper
        \marginparwidth 0.0in
        \parskip 5pt plus 1pt   \jot = 1.5ex}
\catcode`\@=11
\def\marginnote#1{}

\newcount\hour
\newcount\minute
\newtoks\amorpm
\hour=\time\divide\hour by60
\minute=\time{\multiply\hour by60 \global\advance\minute by-\hour}
\edef\standardtime{{\ifnum\hour<12 \global\amorpm={am}%
        \else\global\amorpm={pm}\advance\hour by-12 \fi
        \ifnum\hour=0 \hour=12 \fi
        \number\hour:\ifnum\minute<10 0\fi\number\minute\the\amorpm}}
\edef\militarytime{\number\hour:\ifnum\minute<10 0\fi\number\minute}

\def\draftlabel#1{{\@bsphack\if@filesw {\let\thepage\relax
   \xdef\@gtempa{\write\@auxout{\string
      \newlabel{#1}{{\@currentlabel}{\thepage}}}}}\@gtempa
   \if@nobreak \ifvmode\nobreak\fi\fi\fi\@esphack}
        \gdef\@eqnlabel{#1}}
\def\@eqnlabel{}
\def\@vacuum{}
\def\draftmarginnote#1{\marginpar{\raggedright\scriptsize\tt#1}}

\def\draftlabel#1{{\@bsphack\if@filesw {\let\thepage\relax
   \xdef\@gtempa{\write\@auxout{\string
      \newlabel{#1}{{\@currentlabel}{\thepage}}}}}\@gtempa
   \if@nobreak \ifvmode\nobreak\fi\fi\fi\@esphack}
        \gdef\@eqnlabel{#1}}
\def\@eqnlabel{}
\def\@vacuum{}
\def\draftmarginnote#1{\marginpar{\raggedright\scriptsize\tt#1}}

\def\draft{\oddsidemargin -.5truein
        \def\@oddfoot{\sl preliminary draft \hfil
        \rm\thepage\hfil\sl\today\quad\militarytime}
        \let\@evenfoot\@oddfoot \overfullrule 3pt
        \let\label=\draftlabel
        \let\marginnote=\draftmarginnote
   \def\@eqnnum{(\theequation)\rlap{\kern\marginparsep\tt\@eqnlabel}%
\global\let\@eqnlabel\@vacuum}  }

\def\numberbysection{\@addtoreset{equation}{section}
        \def\theequation{\thesection.\arabic{equation}}}

\def\underline#1{\relax\ifmmode\@@underline#1\else
        $\@@underline{\hbox{#1}}$\relax\fi}

\def\titlepage{\@restonecolfalse\if@twocolumn\@restonecoltrue\onecolumn
     \else \newpage \fi \thispagestyle{empty}\c@page\z@
        \def\thefootnote{\fnsymbol{footnote}} }

\def\endtitlepage{\if@restonecol\twocolumn \else  \fi
        \def\thefootnote{\arabic{footnote}}
        \setcounter{footnote}{0}}  %\c@footnote\z@ }
\relax

\numberbysection
\hybrid

\newfont{\Bbb}{msbm10 scaled 1\@ptsize00}
\newfont{\Bbbb}{msbm7 scaled 1\@ptsize00}
\newcommand{\CC}{\mbox{\Bbb C}}

\newcommand{\DDD}{\raise-1pt\hbox{$\mbox{\Bbbb D}$}}

\newcommand{\UUU}{\raise-1pt\hbox{$\mbox{\Bbbb U}$}}

\newcommand{\ZZ}{\mbox{\Bbb Z}}
\newcommand{\z}{\raise-1pt\hbox{$\mbox{\Bbbb Z}$}}
\newcommand{\cD}{\hat{\mathcal D}}

\def\beq{\begin{equation}}
\def\eeq{\end{equation}}
\def\p{\partial}

\def\G{\Gamma}
\def\LL{\mathcal L}

\newtheorem{theorem}{Theorem}[section]
\newtheorem{lemma}{Lemma}[section]
\newtheorem{lemma-definition}{Lemma-Definition}[section]
\newtheorem{corollary}{Corollary}[section]

\newtheorem{definition}{Definition}[section]
\newtheorem{proposition}{Proposition}[section]

\def\res{\mathop{\hbox{res}}\limits}

\def\square{\hfill
{\vrule height6pt width6pt depth1pt} \break \vspace{.01cm}}

\begin{document}

\begin{titlepage}

\title{Kadomtsev-Petviashvili turning points and CKP hierarchy}

%\title{Quasi-periodic and elliptic solutions to the CKP hierarchy}

\author{I. Krichever\thanks{Skolkovo Institute of Science and
Technology, 143026, Moscow, Russia and
National Research University Higher School of
Economics,
20 Myasnitskaya Ulitsa, Moscow 101000, Russia, and
Columbia University, New York, USA;
e-mail: krichev@math.columbia.edu}
\and
A.~Zabrodin\thanks{
National Research University Higher School of Economics,
20 Myasnitskaya Ulitsa,
Moscow 101000, Russia and
Skolkovo Institute of Science and Technology, 143026, Moscow, Russia and
ITEP NRC KI, 25
B.Cheremushkinskaya, Moscow 117218, Russia;
e-mail: zabrodin@itep.ru}}

\date{December 2020}
\maketitle

\vspace{-7cm} \centerline{ \hfill ITEP-TH-24/20}\vspace{7cm}

\begin{abstract}

A characterization of the Kadomtsev-Petviashvili hierarchy of type C (CKP)
in terms of the KP tau-function is given. Namely, we prove that the
CKP hierarchy can be identified with the restriction of odd times flows
of the KP hierarchy on the locus of turning points of the second flow.
The notion of CKP tau-function is clarified and connected with the KP tau function.
Algebraic-geometrical solutions and in particular elliptic solutions
are discussed in detail. The new identity for theta-functions of
curves with holomorphic involution having fixed points is obtained.
\end{abstract}

\end{titlepage}

\vspace{5mm}

%

%\newpage
\tableofcontents

\vspace{5mm}

\section{Introduction}

The Kadomtsev-Petviashvily (KP) hierarchy is one of the most fundamental in the modern
theory of integrable systems. It has at least three well-known
de\-fi\-ni\-ti\-ons/rep\-re\-sen\-ta\-ti\-ons. In its original,
the so-called Zakharov-Shabat form \cite{ZS75}, it is an infinite
system of equations on an infinite number of variables which are
the coefficients of monic ordinary linear differential operators
\beq\label{Bk}
B_k=\p_x^k+\sum_{i=0}^{k-2} u_{k,i}(x,{\bf t})\p_x^i
\eeq
depending on $x$ and an infinite set of ``times'' ${\bf t}=\{t_1, t_2, t_3, \ldots\}$.
The equations of the hierarchy are equivalent to the operator equations
\beq\label{kp10}
\p_{t_l}B_k-\p_{t_k}B_l+[B_k, B_l]=0, \quad \mbox{for all pairs $k,l$}.
\eeq
For each pair $(k,l)$ the operator equation (\ref{kp10})
is equivalent to a system of partial differential equations on
the coefficients of the operators $B_k,B_l$. The system is well-defined
in the sense that the number of equations is equal to the number of unknown functions.
For example, for the case $k=2, l=3$ in which $B_2=\p^2_x+2u$ and
$B_3=\p_x^3+3u\p_x +w$ equation (\ref{kp10}) is equivalent
to a system of two equations for $u$ and $w$. After eliminating
$w$ from this system, and after the change of the notation for
independent variables $t_2=y, t_3=t$, the remaining equation for $u$
becomes the original KP equation
\beq\label{kporig}
3u_{yy}=\left( 4u_{t}-12u u_x -u_{xxx}\right)_x.
\eeq

\medskip
\noindent
{\bf Remark.}
Under the assumption that $u$ is a periodic function of the variables $x,y$, i.e. $u(x+\ell_1,y)=u(x,y+\ell_2)=u(x,y)$,
the KP hierarchy can be defined as a set of commuting flows on the space
of Cauchy data for (\ref{kporig}): the space of {\it one} periodic function 
of two variables \cite{kp1}.

\medskip

The second form of the KP hierarchy (which is often called the Sato form)
was introduced in \cite{sato} as a system of commuting flows
on the space of sequences $(u_1(x),u_2(x),\ldots) $
of functions of one variable $x$, which can be identified
with the space of pseudo-differential operators of the form
\beq\label{satokp}
\LL=\p_x +u_1\p_x^{-1}+u_2 \p_x^{-2}+\ldots
\eeq
The flows are defined by the Lax equations
\beq\label{kp3}
\p_{t_k} \LL=[B_k, \, \LL], \quad B_k =
\Bigl (\LL^k\Bigr )_+, \quad k=1,2,3, \ldots
\eeq
where $(\cdot)_+$ stands for the differential part of a pseudo-differential operator.
The statement that equations (\ref{kp10}) follow from equations (\ref{kp3}) is easy.
The inverse statement is true up to a triangular change of time variables \cite{takebe}.

The third form of the KP hierarchy is an infinite system of equations for {\it one} function
$\tau^{\rm KP}({\bf t})$ of an {\it infinite} number of variables
generated by the Hirota bilinear equation \cite{JimboMiwa,DJKM83}
\beq\label{ch02}
\oint_{C_\infty}\tau^{\rm KP}({\bf t}-[z^{-1}])\tau^{\rm KP}({\bf t'}+[z^{-1}])\,
\exp \Bigl (\sum_{k\geq 1}(t_k-t_k')z^k\Bigr )dz=0,
\eeq
which should be valid for all sets of times ${\bf t}$, ${\bf t'}$.
Here and below $C_\infty$ is a big circle around the infinity $z=\infty$ and
${\bf t}\pm [z^{-1}]$ denotes the following special shift of time variables:
\beq\label{tshift}
{\bf t}\pm [z^{-1}]:=
\Bigl \{ t_1\pm \frac{1}{z}, t_2\pm \frac{1}{2z^2}, t_3\pm
\frac{1}{3z^3}, \ldots \Bigr \}.
\eeq

Note that equation (\ref{ch02}) is invariant under the transformation
\beq\label{linear}
\tau^{\rm KP}(x, {\bf t}) \longrightarrow \exp \Bigl (\gamma_0 +\gamma_1 x+
\sum_{j\geq 1}\gamma_j t_j\Bigr )\, \tau^{\rm KP}(x, {\bf t})
\eeq
with constant $\gamma_j$. The transformed
tau-function corresponds to the same solution of the KP equations
in the Sato or Zakharov-Shabat form. The tau-functions which differ by multiplication by exponent of a linear function of times
are called equivalent.

In \cite{DJKM81} an infinite integrable hierarchy of partial differential
equations with ${\rm Sp}\, (\infty)$ symmetry was introduced and  called the
Kadomtsev-Petviashvili hierarchy of type C (CKP). It is a hierarchy of commuting flows that are the restriction of the flows of the KP hierarchy corresponding
to ``odd'' times ${\bf t}_{\rm o}=\{t_1, t_3, t_5, \ldots \}$
onto the space of {\it anti self-adjoint}
pseudo-differential operators $\mathcal L_{\rm CKP}$ of the form (\ref{satokp}), i.e. such that
\beq\label{ckp01}
\LL_{\rm CKP}^{\dag}=-\LL_{\rm CKP},
\eeq
where $\dag$ means the formal adjoint
defined by the rule $\Bigl (f(x)\circ \p_x^{m}\Bigr )^{\dag}=(-\p_x)^m \circ f(x)$.
The CKP hierarchy was revisited in \cite{DM-H09,CW13,CH14,LOS12}.

The first goal of this work is to characterize the CKP hierarchy in terms of the
KP tau-function. More precisely, each solution of the CKP
hierarchy has a unique extension to the solution of the full KP
hierachy via the flows (\ref{kp3}) with even $k$ (which obviously do not preserve constraint (\ref{ckp01})). In what follows we will refer to
the corresponding solution to the KP hierarchy as {\it KP extension
of the solution to the CKP hierarchy}.
In section 2 we prove that the KP tau-function is the tau-function of such a solution if the equation
\beq\label{int3}
\p_{t_2}\log \tau^{\rm KP}\Bigl |_{{\bf t}_{\rm e}=0}=0
%=\p_{t_4}\log \tau^{\rm KP}\Bigl |_{{\bf t}_{\rm e}=0}
%=\p_{t_6}\log \tau^{\rm KP}\Bigl |_{{\bf t}_{\rm e}=0}
%=\ldots =0
\eeq
holds for all ${\bf t}_{\rm o}$, where
all ``even'' times ${\bf t}_{\rm e}=(t_2,t_4,\ldots)$ are set equal to zero.
Conversely, in the equivalence class of any KP tau-function which is the tau-function of
the KP extension of a solution to the CKP hierarchy there exists one which satisfies
the condition (\ref{int3}). We note that this condition selects ``turning points''
of the KP hierarchy in the sense that if $x_i$ are zeros of the tau-function
$\tau^{\rm KP}(x, {\bf t})$, i.e., $\tau^{\rm KP}(x_i, {\bf t})=0$, then
$\p_{t_2}x_i(t_1, t_3, \ldots )=0$ for all $t_1, t_3, t_5, \ldots$.

We also prove the existence of the tau-function
$\tau (x,{\bf t}_{\rm o})$ for the
CKP hierarchy which is a function of ``odd'' times ${\bf t}_{\rm o}$ only and
prove that it is the square root of $\tau^{\rm KP}$
satisfying the condition (\ref{int3}):
\beq\label{int3a}
\tau (t_1, t_3, t_5, \ldots , )=\sqrt{\vphantom{A^a}\tau^{\rm KP}(t_1, 0, t_3, 0, t_5, 0,
\ldots )}.
\eeq

In the first part of Section 3 we present in detail the algebraic-geometrical
construction of quasi-periodic solutions to the CKP hierarchy briefly outlined in \cite{DJKM81}.
We start from the general scheme proposed in \cite{Krichever77,Krichever77a}.
The specialization for the CKP hierarchy is a certain reduction of this general scheme.
The data defining the algebraic-geometrical solutions of the CKP hierarchy are the following:
a smooth algebraic curve $\Gamma$ of genus $g$ with a holomorphic involution
having at least one fixed point $P_{\infty}\in \Gamma$, a local parameter
in a neighborhood of $P_{\infty}$ which is
{\it odd} with respect to the involution and a generic {\it admissible}
divisor of degree $g$. The locus of the admissible divisors
in the Jacobian is a translate of the Prym variety of $\Gamma$.
In the second part of Section 3 we prove the new identity for the
Riemann theta-function of a curve with involution having at least
one fixed point (Theorem \ref{thm-main}). The identity is an
algebraic-geometrical incarnation of the relations between KP-
and CKP-tau-functions discussed in Section 2.

In Section 4 we study double-periodic (elliptic) in the variable $x=t_1$ solutions to the
$C$-version of the KP equation and their trigonometric and rational degenerations.

In the seminal paper \cite{AMM77} the motion of poles of singular solutions to the
Korteweg-de Vries and Boussinesq equations was considered. It was discovered that the poles move as particles of the many-body Calogero-Moser system \cite{Calogero71,Calogero75,Moser75}
with some additional restrictions in the phase space.
In \cite{Krichever78,CC77} it was shown that in the case of the KP
equation this correspondence becomes an isomorphism: the dynamics of
poles of rational solutions to the
KP equation is given by equations of motion for the Calogero-Moser system
with pairwise interaction potential
$1/(x_i-x_j)^2$. This remarkable connection
was further generalized to elliptic
solutions in \cite{Krichever80}:
poles $x_i$ of the elliptic solutions as functions
of $t_2=y$ move according to the equations of motion
\beq\label{int1}
\p_y^2 x_i=4\sum_{k\neq i} \wp ' (x_i-x_k)
\eeq
of Calogero-Moser particles with the elliptic
interaction potential $\wp (x_i-x_j)$ ($\wp$ is the Weierstrass $\wp$-function).
Moreover, in \cite{Krichever80} it was shown that the origin of
equations (\ref{int1}) is related to a more fundamental problem:
when a linear equation with elliptic coefficients has {\it double-Bloch} solutions
(i.e. solution which are sections of a line bundle over
the elliptic curve, see \cite{kr-nested}).
Recently, the method proposed in \cite{Krichever80} was applied
to the theory of elliptic solutions of the BKP equation \cite{RZ20,Z19}.

Along the same line of arguments we derive in Section 4 the equations
of motion for poles of elliptic solutions to the CKP equation:
\beq\label{int2}
\dot x_i =3\sum_{k\neq i}^n\wp (x_i-x_k)-6c,
\eeq
where $c$ is a constant and dot means the $t_3$-derivative.
In contrast to the KP and BKP cases, where
the equations of motion are of the second order
(see \cite{Krichever80,RZ20,Z19}) equations (\ref{int2}) are of the first order.
As follows from the comparison of the CKP and KP hierarchies in
Section 2 equation (\ref{int2}) coincide with the restriction of the Calogero-Moser
flow corresponding to the higher Hamiltonian $H_3$
to the manifold of {\it turning points} in the $2n$-dimensional
phase space $(p_i, x_i)$, i.e. the $n$-dimensional submanifold
$p_i=\p_y x_i=0$ for all $i=1, \ldots , n$.

\medskip\noindent
{\bf Remark.}
The notion of the turning points of the elliptic
Calogero-Moser system and the study of the corresponding spectral
curves in the forthcoming paper \cite{KN} was motivated by the
problem of construction of explicit solutions to the
two-dimensional $O(2m+1)$ sigma-model.

\medskip

\section{The CKP hierarchy}

\subsection{The $\LL$-operator and the dressing operator}

The set of independent variables (``times'') in the CKP hierarchy is
${\bf t}_{\rm o}=\{t_1, t_3, t_5, \ldots \}$.
Like in the BKP case, they are indexed by positive odd numbers. It is convenient
to set $t_1 = x+\mbox{const}$, so that the vector fields $\p_{t_1}$ and $\p_x$ are
identical: $\p_{t_1}=\p_x$.
The hierarchy is defined on the space of pseudo-differential operators $\LL_{\rm CKP}$
of the form
\beq\label{ckp1}
\LL_{\rm CKP}=\p_x +u_1\p_x^{-1}+u_2 \p_x^{-2}+\ldots
\eeq
subject to the constraint
\beq\label{ckp2}
\LL_{\rm CKP}^{\dag}=-\LL_{\rm CKP},
\eeq
The coefficients $u_j$ of $\LL_{\rm CKP}$ depend on $x$ and on all the times. 
It is convenient to introduce the wave operator (or dressing operator)
\beq\label{ckp1a}
W=1+\xi_1 \p_x^{-1}+\xi_2 \p_x^{-2}+\ldots
\eeq
such that
\beq\label{ckp1ab}
\LL_{\rm CKP}=W\p_x W^{-1}.
\eeq
The wave operator is unique up to multiplication from the right by a pseudo-differential operator with constant coefficients.

The constraint (\ref{ckp2}) implies that $W^{\dag}W$ commutes with $\p_x$, i.e.,
it is a pseudo-differential operator with constant coefficients. 
We fix the above mentioned ambiguity in the definition of the 
wave operator by imposing the equation
$W^{\dag}W=1$, i.e.
\beq\label{ckp1b}
W^{\dag}=W^{-1}.
\eeq

The hierachy of flows is defined by the Lax equations
\beq\label{ckp3}
\p_{t_k}\LL_{\rm CKP}=[B_k, \, \LL_{\rm CKP}], \quad B_k =
\Bigl (\LL_{\rm CKP}^k\Bigr )_+, \quad k=1,3,5, \ldots ,
\eeq
which obviously preserve the constraint (\ref{ckp1b}) since $B_k^{\dag}=-B_k$ (for odd $k$).

The zero curvature (Zakharov-Shabat) equations
\beq\label{ckp3a}
\p_{t_l}B_k-\p_{t_k}B_l+[B_k, B_l]=0, \quad \mbox{$k,l$ odd}
\eeq
is an easy corollary of (\ref{ckp3}). They are equivalent to the statement that the flows (\ref{ckp3}) commute with each other.

The first equation of the CKP hierarchy is the equation
$\p_{t_3}B_5-\p_{t_5}B_3+[B_5, B_3]=0$ with
\beq\label{ckp5}
\begin{array}{c}
B_3 =\p_x^3 +6u\p_x +3u', \quad u'\equiv \p_x u, \quad u=\frac{1}{2}\, u_1,
\end{array}
\eeq
\beq\label{B5}
\begin{array}{c}
B_5 =\p_x^5 +10u\p^3_x +15u' \p_x^2 +v\p_x +\frac{1}{2}\, (v'-5u''').
\end{array}
\eeq
Straightforward calculations give the following
system of equations for the unknown functions $u,v$:
\beq\label{ckp0}
\left \{ \begin{array}{l}
10\p_{t_3}u=3v' -35u''' -120 uu'
\\ \\
6\p_{t_5}u-\p_{t_3}v=\frac{57}{2}\, u''''' +150 uu''' +180u'u'' -
\frac{5}{2}\, v''' +6vu'-6uv'.
\end{array} \right.
\eeq
Note that the variable $v$ can be excluded by passing to the unknown function $U$
such that $U'=u$.

\subsection{The wave function and the tau-function}

\label{section:wave}

The Lax equations (\ref{ckp3}) are compatibility conditions of the auxiliary linear problems
\beq\label{ckp4}
\p_{t_k}\Psi =B_k \Psi , \quad \LL_{\rm  CKP}\Psi =z\Psi
\eeq
for the formal wave function
\beq\label{ckp4a}
\Psi=\Psi({\bf t}_{\rm o}, z)=We^{xz+\zeta ({\bf t}_{\rm o}, z)},
\eeq
where $z$ is the spectral parameter and
\beq\label{ckp7}
\zeta ({\bf t}_{\rm o}, z)=\sum_{k\geq 1, \, {\rm odd}}t_k z^k.
\eeq
Note that the operator $\p_x^{-1}$ acts to the exponential function as $\p_x^{-1}e^{xz}=z^{-1}e^{xz}$. Therefore, from (\ref{ckp1a}), (\ref{ckp4a}), it follows that
the wave function  has the following expansion as $z\to \infty$:
\beq\label{ckp5b}
\Psi(x, {\bf t}_{\rm o}, z)=e^{xz+\zeta ({\bf t}_{\rm o}, z)}
\Bigl (1+\sum_{k\geq 1}\xi_k z^{-k}\Bigr ).
\eeq

\begin{proposition} (\cite{DJKM81})
The wave function $\Psi$ satisfies the
bilinear relation
\beq\label{ckp5a}
\oint_{C_{\infty}}\! \Psi (x, {\bf t}_{\rm o}, z)\Psi (x, {\bf t}_{\rm o}', -z){dz}=0
\eeq
for all ${\bf t}_{\rm o}, {\bf t}_{\rm o}'$.
\end{proposition}
\noindent
For completeness, we give a sketch of the proof here.
By virtue of differential equations (\ref{ckp3}), the bilinear relation is equivalent
to vanishing of the coefficients
$$
b_m=\frac{1}{2\pi i}\,
\p_{x'}^m \oint_{C_{\infty}}\! \Psi (x, {\bf t}_{\rm o}, z)\Psi (x', {\bf t}_{\rm o}, -z)
{dz} \Biggr |_{x'=x}\quad \mbox{for all $m\geq 0$.}
$$
We have:
$$
b_m=\frac{1}{2\pi i}\oint_{C_{\infty}}\Bigl (\sum_{k\geq 0}\xi_k(x)z^{-k}\Bigr )\p_{x'}^m
\Bigl (\sum_{l\geq 0}\xi_l(x')(-z)^{-l}\Bigr )e^{(x-x')z} {dz} \Biggr |_{x'=x}
$$
$$
=\frac{1}{2\pi i}\oint_{C_{\infty}}\Bigl (\sum_{k\geq 0}\xi_kz^{-k}\Bigr )(\p_x -z)^m
\Bigl (\sum_{l\geq 0}\xi_l(-z)^{-l}\Bigr ){dz}
$$
$$
=\sum_{j+k+l=m+1}(-1)^{m+j+l}\left (\! \begin{array}{c}m\\ j \end{array} \! \right )
\xi_k \p_x^j \xi_l.
$$
The last expression is the coefficient at $(-1)^m \p_x^{-m-1}$ in the operator
$WW^{\dag}$:
$$
WW^{\dag}=1+\sum_{m\geq 0}(-1)^m b_m \p_x^{-m-1}.
$$
Since $WW^{\dag}=1$ (see (\ref{ckp1b})), we get that $b_m=0$ for all $m\geq 0$.

\begin{theorem}
\label{theorem:exist}
There exists a function $\tau = \tau (x, {\bf t}_{\rm o})$
such that
\beq\label{e4}
\Psi (x, {\bf t}_{\rm o},z)  =
%z^{-1/2} \sqrt{\vphantom{B^{a^a}}\p_x \log \psi} \cdot \psi =
(2z)^{-1/2}\sqrt{\vphantom{B^{a^a}}\p_x \psi^2(x, {\bf t}_{\rm o},z)},
\eeq
where
\beq\label{e3}
\psi (x, {\bf t}_{\rm o},z) = e^{xz+\zeta ({\bf t}_{\rm o}, z)}
\frac{\tau (x, {\bf t}_{\rm o}-2[z^{-1}]_{\rm o})}{\tau (x, {\bf t}_{\rm o})}
\eeq
and we use the notation
\beq\label{ckp8}
{\bf t}_{\rm o}+j[z^{-1}]_{\rm o}:= \Bigl \{ t_1 +\frac{j}{z},  t_3 +\frac{j}{3z^3},
t_5 +\frac{j}{5z^5}, \, \ldots \Bigr \}, \quad j\in \ZZ .
\eeq
\end{theorem}
\begin{definition}The function $\tau = \tau (x, {\bf t}_{\rm o})$ is called the tau-function
of the CKP hierarchy.
\end{definition}

\noindent
{\it Proof of Theorem \ref{theorem:exist}.}
Representing the right hand side of (\ref{e4}) in the explicit form,
we see that we should prove the formula
\beq\label{ckp6}
\Psi  = e^{xz+\zeta ({\bf t}_{\rm o}, z)}G(x, {\bf t}_{\rm o}, z)\,
\frac{\tau (x, {\bf t}_{\rm o}-2[z^{-1}]_{\rm o})}{\tau (x, {\bf t}_{\rm o})},
\eeq
where
\beq\label{ckp6a}
G(x, {\bf t}, z)=\left (1+z^{-1}\p_{t_1}\log \frac{\tau (x, {\bf t}_{\rm o}
-2[z^{-1}]_{\rm o})}{\tau (x,{\bf t}_{\rm o})}
\right )^{1/2}.
\eeq
The proof is based on the bilinear relation (\ref{ckp5a}).
Let us represent the wave function in the form
$$
\Psi (x, {\bf t}_{\rm o}, z)=e^{xz+\zeta ({\bf t}_{\rm o}, z)}w(x,{\bf t}_{\rm o}, z)
$$
and set ${\bf t}_{\rm o}'={\bf t}_{\rm o}-2[a^{-1}]_{\rm o}$ in the bilinear relation. We have
$\displaystyle{e^{\zeta(2[a^{-1}]_{\rm o}, z)}=
\frac{a+z}{a-z}}$. Then the residue calculus yields
\beq\label{der1}
w({\bf t}_{\rm o}, a) w({\bf t}_{\rm o}-2[a^{-1}]_{\rm o}, -a)=f({\bf t}_{\rm o}, a),
\eeq
where
\beq\label{der2}
f({\bf t}_{\rm o}, z)=1+\frac{1}{2z}\, \Bigl (\xi_1({\bf t}_{\rm o})
-\xi_1({\bf t}_{\rm o}-2[z^{-1}]_{\rm o})\Bigr )
\eeq
and we do not indicate the dependence on $x$ for brevity.
Next, we set ${\bf t}_{\rm o}'={\bf t}_{\rm o}-2[a^{-1}]_{\rm o}-2[b^{-1}]_{\rm o}$
in the bilinear relation and the residue calculus yields
\beq\label{der7}
\begin{array}{c}
\displaystyle{\frac{a+b}{a-b}\Bigl (aw({\bf t}_{\rm o}, a)w({\bf t}_{\rm o}\! -\!
2[a^{-1}]_{\rm o}\!
-\! 2[b^{-1}]_{\rm o}, -a)-
bw({\bf t}_{\rm o}, b)w({\bf t}_{\rm o}\! -\! 2[a^{-1}]_{\rm o}
\! -\! 2[b^{-1}]_{\rm o}, -b)\Bigr )}
\\ \\
\displaystyle{=
a+b+\frac{1}{2}\Bigl (\xi_1({\bf t}_{\rm o})-\xi_1({\bf t}_{\rm o}\! -\! 2[a^{-1}]_{\rm o}
\! -\! 2[b^{-1}]_{\rm o}\Bigr )}.
\end{array}
\eeq
Using the relation (\ref{der1}), we can represent this equation in the form
\beq\label{der8}
\begin{array}{c}
\displaystyle{
\frac{1}{a-b}\left (af({\bf t}_{\rm o}\! -\! 2[b^{-1}]_{\rm o}, a)
\frac{w({\bf t}_{\rm o}, a)}{w({\bf t}_{\rm o}\! -\! 2[b^{-1}]_{\rm o},a)}-
bf({\bf t}_{\rm o}\! -\! 2[a^{-1}]_{\rm o}, b)
\frac{w({\bf t}_{\rm o}, b)}{w({\bf t}_{\rm o}\! -\! 2[a^{-1}]_{\rm o},b)}\right )
}
\\ \\
\displaystyle{=\, 1+\frac{\xi_1({\bf t}_{\rm o})-
\xi_1({\bf t}_{\rm o}\! -\! 2[a^{-1}]_{\rm o}\! -\! 2[b^{-1}]_{\rm o})}{2(a+b)}.}
\end{array}
\eeq
Shifting here ${\bf t}_{\rm o}\rightarrow {\bf t}_{\rm o}+2[b^{-1}]_{\rm o}$,
changing the sign of $b$ (i.e, changing $b\to -b$)
and using (\ref{der1}) in the second term in the
left hand side after that, we arrive at the equation
\beq\label{der9}
\begin{array}{c}
\displaystyle{
\frac{1}{a+b}\left (af({\bf t}_{\rm o}, a)
\frac{w({\bf t}_{\rm o}\! -\! 2[b^{-1}]_{\rm o}, a)}{w({\bf t}_{\rm o},a)}-
bf({\bf t}_{\rm o}, b)
\frac{w({\bf t}_{\rm o}\! -\! 2[a^{-1}]_{\rm o}, b)}{w({\bf t}_{\rm o},b)}\right )
}
\\ \\
\displaystyle{=\, 1+\frac{\xi_1({\bf t}_{\rm o}\! -\! 2[b^{-1}]_{\rm o})-
\xi_1({\bf t}_{\rm o}\! -\! 2[a^{-1}]_{\rm o})}{2(a-b)}.}
\end{array}
\eeq
Together equations (\ref{der8}), (\ref{der9}) form the system of equations
\beq\label{der10}
\left \{
\begin{array}{l}
\displaystyle{\frac{1}{a\! -\! b}\left (af({\bf t}_{\rm o}\! -\! 2[b^{-1}]_{\rm o}, a)X_a^{-1}\! -\!
bf({\bf t}_{\rm o}\! -\! 2[a^{-1}]_{\rm o}, b)X_b^{-1}\right )\! =\!
\frac{af({\bf t}_{\rm o}, a)\! +\! bf({\bf t}_{\rm o}\! -\! 2[a^{-1}]_{\rm o}, b)}{a+b}}
\\ \\
\displaystyle{\frac{1}{a\! +\! b}\, \Bigl (af({\bf t}_{\rm o}, a)X_a -
bf({\bf t}_{\rm o}, b)X_b\Bigr ) =
\frac{af({\bf t}_{\rm o}, a) - bf({\bf t}_{\rm o}, b)}{a-b}}
\end{array}
\right.
\eeq
for the ``unknowns''
\beq\label{der11}
X_a=\frac{w({\bf t}_{\rm o}-2[b^{-1}]_{\rm o}, a)}{w({\bf t}_{\rm o}, a)}, \qquad
X_b=\frac{w({\bf t}_{\rm o}-2[a^{-1}]_{\rm o}, b)}{w({\bf t}_{\rm o}, b)}.
\eeq
Multiplying the two equations (\ref{der10}), one obtains, using
the identity
\beq\label{der12}
af({\bf t}_{\rm o}, a)-af({\bf t}_{\rm o}-2[b^{-1}]_{\rm o}, a)
-bf({\bf t}_{\rm o}, b)+bf({\bf t}_{\rm o}-2[a^{-1}]_{\rm o}, b)=0,
\eeq
the following simple relation:
\beq\label{der13}
\frac{w({\bf t}_{\rm o}, a)
w({\bf t}_{\rm o}-2[a^{-1}]_{\rm o}, b)}{w({\bf t}_{\rm o}, b)
w({\bf t}_{\rm o}-2[b^{-1}]_{\rm o}, a)}=
\left (\frac{f({\bf t}_{\rm o}, a)
f({\bf t}_{\rm o}-2[a^{-1}]_{\rm o}, b)}{f({\bf t}_{\rm o}, b)
f({\bf t}_{\rm o}-2[b^{-1}]_{\rm o}, a)}\right )^{1/2}.
\eeq
Therefore, introducing $w_0({\bf t}_{\rm o}, z)=w({\bf t}_{\rm o}, z)
f^{-1/2}({\bf t}_{\rm o}, z)$, we get
\beq\label{der14}
\frac{w_0({\bf t}_{\rm o}, a)
w_0({\bf t}_{\rm o}-2[a^{-1}]_{\rm o}, b)}{w_0({\bf t}_{\rm o}, b)
w_0({\bf t}_{\rm o}-2[b^{-1}]_{\rm o}, a)}=1.
\eeq

Our goal is to prove that there exists a function $\tau ({\bf t}_{\rm o})$ such that
\beq\label{der15}
w_0({\bf t}_{\rm o}, z)=\frac{\tau ({\bf t}_{\rm o}-2[z^{-1}]_{\rm o})}{\tau ({\bf t}_{\rm o})}.
\eeq
For that it is enough to show that there is a function $\tau$ such that the equation
\beq\label{der151}
\cD (w_0({\bf t}_{\rm o}, z)+\tau ({\bf t}_{\rm o}))=0
\eeq
with
\beq\label{cD}
\cD:=\p_z-2\! \sum_{m\geq 1, \,\, {\rm odd}}\! z^{-m-1}\p_{t_m}
\eeq
holds.

Indeed, integrating  equation $\cD F=0$ along its characteristics we get that a function
$F({\bf t}_{\rm o}, z)$ is in the kernel of the differential operator $\cD$ if and only it is of the form
$$F({\bf t}_{\rm o}, z)=f ({\bf t}_{\rm o}-2[z^{-1}]_{\rm o}) $$
for some function $f({\bf t}_{\rm o})$. For $F$ as in (\ref{der151}) the initial condition $w_0({\bf t}_{\rm o}, \infty)=0$ allows to identify the corresponding function $f$ with $\tau$.

Equation (\ref{der151}) is equivalent to the equations
$$
Y_n :=\res_{z=\infty} \Bigl [ z^n \cD \log w_0 \Bigr ]=2\frac{\p \log \tau}{\p t_n}.
$$
Therefore, to complete the proof of the existence of the tau-function it remains only to show that
$\p_{t_n}Y_m({\bf t}_{\rm o})=\p_{t_m}Y_n({\bf t}_{\rm o})$.

Changing $a\to z$, $b\to \zeta$ in (\ref{der14}), and applying the operator
$\cD$ to logarithm of this equality, we get
$$
\cD\left(\log w_0({\bf t}_{\rm o}, z)-\log w_0({\bf t}_{\rm o}\! -\! 2[\zeta ^{-1}]_{\rm o}, z)
+\log w_0({\bf t}_{\rm o}, \zeta )\right)=0,
$$
or
\beq\label{der16}
Y_n ({\bf t}_{\rm o})-Y_n ({\bf t}_{\rm o}-2[\zeta^{-1}]_{\rm o})=-2
\p_{t_n}\log w_0({\bf t}_{\rm o}, \zeta ).
\eeq
Denote $F_{mn}=\p_{t_m}Y_n-\p_{t_n}Y_m$. Then, from (\ref{der16}) it follows
that the equation
\beq\label{der17}
F_{mn}({\bf t}_{\rm o})=F_{mn}({\bf t}_{\rm o}-2[\zeta^{-1}]_{\rm o})
\eeq
holds identically in $\zeta$. Expanding the right hand side in a power series,
$$
\begin{array}{c}
F_{mn}({\bf t}_{\rm o}\! -\! 2[\zeta^{-1}]_{\rm o})=F_{mn}({\bf t}_{\rm o})
\! -\! 2\zeta^{-1}\p_{t_1}F_{mn}({\bf t}_{\rm o})\! -\!  \frac{2}{3}\, \zeta^{-3}(\p_{t_3}
F_{mn}({\bf t}_{\rm o})\! +\! 2 \p_{t_1}^3F_{mn}({\bf t}_{\rm o}))+\ldots ,
\end{array}
$$
we see from the $\zeta^{-1}$-term that $F_{mn}$ does not depend on $t_1$. Then from
the $\zeta^{-3}$-term we conclude that it does not depend on $t_3$ and so on, so it
does not depend on $t_k$ for all (odd) $k$:
$F_{mn}=2a_{mn}$, where $a_{mn}$ are some constants such that $a_{mn}=-a_{nm}$.
Therefore, we can write
$$
Y_n =\sum_m a_{mn}t_m +\p_{t_n}h,
$$
where $h=h({\bf t}_{\rm o})$ is some function. Then from (\ref{der16}) we have
$$
-2\p_{t_n}\log w_0({\bf t}_{\rm o}, z)=\p_{t_n}(h({\bf t}_{\rm o})-
h({\bf t}_{\rm o}-2[z^{-1}]_{\rm o}))+2\! \sum_{m\,\, {\rm odd}}\frac{a_{mn}}{m}\, z^{-m},
$$
or, after integration,
$$
\log w_0({\bf t}_{\rm o}, z)=\frac{1}{2}\, h({\bf t}_{\rm o}\! -\!
2[z^{-1}]_{\rm o}) -\frac{1}{2}\, h({\bf t}_{\rm o})-
\sum_{m\,\, {\rm odd}}\frac{a_{mn}}{m}\, z^{-m}t_n
+\varphi (z),
$$
where $\varphi (z)$ is a function of $z$ only. Substituting this into logarithm of
(\ref{der14}), we conclude that $a_{mn}=0$.

Now, writing $w({\bf t}_{\rm o}, z)=f^{1/2}({\bf t}_{\rm o}, z)w_0({\bf t}_{\rm o}, z)$
and noting that $f({\bf t}_{\rm o}, z)=1+O(z^{-2})$, we see that
\beq\label{der18}
\xi_1 ({\bf t}_{\rm o})=-2\p_{t_1}\log \tau ({\bf t}_{\rm o})
\eeq
and we arrive at (\ref{ckp6})
with $G=f^{1/2}$.
\square

\medskip\noindent
{\bf Remark.} The proof given above is rather involved.
It is instructive to obtain (\ref{ckp6}) up to a common
$x$-independent factor in the following easy way \cite{DM-H09,CW13}.
Let us apply $\p_{t_1}$ to
(\ref{ckp5a}) and set ${\bf t}_{\rm o}'={\bf t}_{\rm o}-2[a^{-1}]_{\rm o}$.
The residue calculus yields
\beq\label{der3}
\begin{array}{c}
2a^2\Bigl (1\! -\! w({\bf t}_{\rm o}, a) w({\bf t}_{\rm o}\! -\! 2[a^{-1}]_{\rm o}, -a)\Bigr )-2a
w'({\bf t}_{\rm o}, a) w({\bf t}_{\rm o}-2[a^{-1}]_{\rm o}, -a)
\\ \\
+2a\Bigl (\xi_1({\bf t}_{\rm o})-\xi_1({\bf t}_{\rm o}\! -\!
2[a^{-1}]_{\rm o})\Bigr ) +\xi_2({\bf t}_{\rm o}\! -\! 2[a^{-1}]_{\rm o})
+\xi_2({\bf t}_{\rm o})+\xi_1'({\bf t}_{\rm o})
\\ \\
\phantom{aaaaaaaaaaaaaaaaaaaaaaaa}-
\xi_1({\bf t}_{\rm o})\xi_1({\bf t}_{\rm o}\! -\! 2[a^{-1}]_{\rm o})=0,
\end{array}
\eeq
where prime means the $x$-derivative and
we again do not indicate the dependence on $x$ explicitly.
Letting $a\to \infty$, we get the relation
\beq\label{der4}
2\xi_2 ({\bf t}_{\rm o})=\xi_1^2 ({\bf t}_{\rm o})-\xi_1' ({\bf t}_{\rm o})
\eeq
which also directly follows from $WW^{\dag}=1$.
Plugging it back to (\ref{der3}), we can rewrite equation (\ref{der3}) in the form
\beq\label{der5}
\begin{array}{c}
w'({\bf t}_{\rm o}, a) w({\bf t}_{\rm o}-2[a^{-1}]_{\rm o}, -a)=
af({\bf t}_{\rm o}, a)(f({\bf t}_{\rm o}, a)-1)+\frac{1}{2}
f'({\bf t}_{\rm o}, a).
\end{array}
\eeq
Using (\ref{der1}), we conclude that
\beq\label{der6}
\begin{array}{c}
\p_x \log w({\bf t}_{\rm o}, a)=a(f({\bf t}_{\rm o}, a)-1)+\frac{1}{2}\, \p_x
\log f({\bf t}_{\rm o}, a)
\\ \\
=\frac{1}{2}\Bigl (\xi_1({\bf t}_{\rm o})-\xi_1({\bf t}_{\rm o}-2[a^{-1}]_{\rm o})\Bigr )+
\frac{1}{2}\, \p_x \log f({\bf t}_{\rm o}, a).
\end{array}
\eeq
Now, setting $\xi_1(x, {\bf t}_{\rm o})=-2\p_x\log \tau (x, {\bf t}_{\rm o})$ and integrating,
we arrive at (\ref{ckp6})
with $G=f^{1/2}$ up to a common multiplier which does not depend on $x$.

\medskip
\noindent{\bf Remark}. Substitution of (\ref{ckp5b}) into (\ref{ckp4}) with $k=3$ gives that the function $u$ in (\ref{ckp5}) equals
\beq\label{ckp5d}
u=-\frac{1}{2}\, \xi_1'=\p_x^2\log \tau
\eeq

\subsection{CKP hierarchy versus KP hierarchy}

The goal of this section is to prove that the CKP hierarchy can be
identified with the restriction of {\it odd-times} flows
of the KP hierarchy onto the locus of {\it turning points of even-times flows}.

%$$
%{\bf t}+j[z^{-1}]=
%\Bigl \{ t_1+\frac{j}{z}, t_2+\frac{j}{2z^2}, t_3+\frac{j}{3z^3}, \ldots \Bigr \}, \quad
%j\in \ZZ .
%$$

Recall that wave function $\Psi^{\rm KP}$ and the
adjoint wave function $\Psi^{\dag \rm KP}$ of the KP hierarchy are expressed through the
tau-function $\tau^{\rm KP}$ as
\beq\label{ch1}
\Psi^{\rm KP}(x, {\bf t}; z)=
\exp \Bigl (xz+\sum_{k\geq 1}t_k z^k\Bigr )
\frac{\tau^{\rm KP} (x, {\bf t}-[z^{-1}] )}{\tau^{\rm KP} (x, {\bf t})},
\eeq
\beq\label{ch1a}
\Psi^{\dag \rm KP}(x, {\bf t}; z)=
\exp \Bigl (-xz-\sum_{k\geq 1}t_k z^k\Bigr )
\frac{\tau^{\rm KP} (x, {\bf t}+[z^{-1}] )}{\tau^{\rm KP} (x, {\bf t})}.
\eeq
where the notation (\ref{tshift}) for the special shift of times is used.
The origin of these expressions is the bilinear relation \cite{JimboMiwa}
\beq\label{ch2}
\oint_{C_{\infty}}\Psi^{\rm KP}(x, {\bf t}, z)
\Psi^{\dag \rm KP}(x, {\bf t}', z) dz=0
\eeq
equivalent to (\ref{ch02}).

A direct consequence of the bilinear relation (\ref{ch2}) with the wave functions
given by (\ref{ch1}), (\ref{ch1a}) is the Hirota-Miwa equation
for the tau-function of the KP hierarchy
\beq\label{tau6b}
\begin{array}{l}
(z_1-z_2)(z_3-z_4)\tau^{\rm KP} ({\bf t}-[z_1^{-1}]-[z_2^{-1}])
\tau^{\rm KP} ({\bf t}-[z_3^{-1}]-[z_4^{-1}])
\\ \\
\phantom{aaaaa}
+(z_2-z_3)(z_1-z_4)\tau^{\rm KP} ({\bf t}-[z_2^{-1}]-[z_3^{-1}])
\tau^{\rm KP} ({\bf t}-[z_1^{-1}]-[z_4^{-1}])
\\ \\
\phantom{aaaaaaaaaa}
+ (z_3-z_1)(z_2-z_4)\tau^{\rm KP} ({\bf t}-[z_1^{-1}]-[z_3^{-1}])
\tau^{\rm KP} ({\bf t}-[z_2^{-1}]-[z_4^{-1}])=0.
\end{array}
\eeq
It is a generating equation for the differential equations of the hierarchy.
The differential equations are obtained by expanding it in negative powers of
$z_1, z_2, z_3, z_4$.
In the limit $z_4\to \infty$, $z_3\to \infty$ equation (\ref{tau6b}) becomes
\beq\label{ch3}
\begin{array}{l}
\displaystyle{
\p_{x}\log \frac{\tau^{\rm KP}\Bigl (x, {\bf t}+
[z_1^{-1}]-[z_2^{-1}]\Bigr )}{\tau^{\rm KP}(x, {\bf t})}}
\\ \\
\phantom{aaaaaaaaaaaaa}\displaystyle{=
(z_2-z_1)\left (\frac{\tau^{\rm KP}\Bigl (x, {\bf t}+
[z_1^{-1}]\Bigr )\tau^{\rm KP}\Bigl (x, {\bf t}-
[z_2^{-1}]\Bigr )}{\tau^{\rm KP}(x, {\bf t})
\tau^{\rm KP}\Bigl (x, {\bf t}+
[z_1^{-1}]-[z_2^{-1}]\Bigr )}-1\right )}.
\end{array}
\eeq
We will need a particular case of (\ref{ch3}) at $z_2=-z_1=z$ which we write
in the form
\beq\label{ch3a}
\frac{1}{2z}\, \p_x \!\! \left (e^{2xz} \frac{\tau^{\rm KP}
\Bigl (x,{\bf t}+[-z^{-1}]-[z^{-1}]\Bigr )}{\tau^{\rm KP}({x,\bf t})}\right )=
e^{2xz}\frac{\tau^{\rm KP}
\Bigl (x,{\bf t}+[-z^{-1}]\Bigr )\tau^{\rm KP}\Bigl (x,{\bf t}-[z^{-1}]\Bigr )}{(\tau^{\rm KP}({x,\bf t}))^2}.
\eeq

The following theorem gives an expression for the CKP tau-functions in terms of the KP tau-functions satisfying the ``turning points'' constraint (\ref{int3}).

\begin{theorem}
\label{theorem:conditions}
The KP tau-function $\tau^{\rm KP}(x, {\bf t})$
is the KP extension of a solution of the CKP hierarchy if the equation
\beq\label{ch6}
\p_{t_2}\log \tau^{\rm KP}\Bigr |_{{\bf t}_{\rm e}=0}=0
\eeq
%\beq\label{ch6a}
%\p_{t_2}\log \tau^{\rm KP}\Bigr |_{{\bf t}_{\rm e}=0}
%=\p_{t_4}\log \tau^{\rm KP}\Bigr |_{{\bf t}_{\rm e}=0}
%=\p_{t_6}\log \tau^{\rm KP}\Bigr |_{{\bf t}_{\rm e}=0}
%=\ldots =0
%\eeq
holds for all $t_1, t_3, t_5, \ldots$ when ``even'' times
${\bf t}_{\rm e}=\{t_2, t_4, t_6, \ldots \}$ are set equal to zero.
Conversely, in the equivalence class of any KP tau-function
corresponding to KP extension of a solution to the CKP hierarchy there is one
which satisfies (\ref{ch6}).
Moreover, the CKP tau-function defined in Theorem \ref{theorem:exist} is equal to
\beq\label{e1}
\tau (x,{\bf t}_{\rm o})=\sqrt{\tau^{\rm KP}(x,t_1, 0, t_3, 0, \ldots )}.
\eeq
\end{theorem}

\medskip

\noindent
{\it Proof}.
Comparing (\ref{ch2}) and (\ref{ckp5a}), we see that the wave functions of the CKP and KP
hierarchies are related as
$$
\begin{array}{l}
\Psi (x, {\bf t}_{\rm o}, z)=e^{\chi (z)}\Psi^{\rm KP}(x, t_1, 0, t_3, 0, \ldots , z),
\\ \\
\Psi (x, {\bf t}_{\rm o}, -z)=e^{-\chi (z)}
\Psi^{\dag \rm KP}(x, t_1, 0, t_3, 0, \ldots , z)
\end{array}
$$
with some function $\chi (z)$ such that $\chi (\infty )=0$, i.e.
\beq\label{ch4a}
\Psi^{\dag \rm KP}(x, t_1, 0, t_3, 0, \ldots , z)=
e^{2\chi_{\rm e}(z)}\Psi^{\rm KP}(x, t_1, 0, t_3, 0, \ldots , -z),
\eeq
where $\chi_{\rm e}(z)=\frac{1}{2}(\chi(z)+\chi (-z))$ is the even part of the function $\chi (z)$.
From (\ref{ch1}), (\ref{ch1a}) and (\ref{ch4a}) it follows
that the KP tau-function is the extension of a solution of the CKP hierarchy
if and only if the equation
\beq\label{ch4}
\begin{array}{l}
\tau^{\rm KP}\Bigl (x, t_1+z^{-1}, \frac{1}{2}\, z^{-2},
t_3+\frac{1}{3}\, z^{-3}, \frac{1}{4}\, z^{-4}, \ldots \Bigr )
\\ \\
\phantom{aaaaaaaaaaa}=e^{2\chi_{\rm e}(z)}\tau^{\rm KP}\Bigl (x, t_1+z^{-1}, -\frac{1}{2}\, z^{-2},
t_3+\frac{1}{3}\, z^{-3}, -\frac{1}{4}\, z^{-4}, \ldots \Bigr )
\end{array}
\eeq
holds identically for all $z, x, t_1, t_3, t_5, \ldots$. Shifting the odd times, we can
rewrite this condition as
\beq\label{ch4b}
\begin{array}{c}
\log \tau^{\rm KP}\Bigl (x, t_1, \frac{1}{2}\, z^{-2},
t_3, \frac{1}{4}\, z^{-4}, \ldots \Bigr )
-\log \tau^{\rm KP}\Bigl (x, t_1, -\frac{1}{2}\, z^{-2},
t_3, -\frac{1}{4}\, z^{-4}, \ldots \Bigr )=2\chi_{\rm e}(z).
\end{array}
\eeq
Comparing the coefficients at $z^{-2}$ of the expansions of the left and right hand sides of (\ref{ch4b})
and passing to an equivalent tau-function
if necessary, we get (\ref{ch6}), i.e. the ``only if'' part of the theorem statement is proven.

We begin the proof of the ``if part'' by the following lemma.
\begin{lemma}
\label{proposition:even}
On solutions of the KP hierarchy
equation (\ref{ch6}) implies that
all derivatives of odd degree higher then $1$ with respect
to various even times are equal to zero for all $x, {\bf t}_{\rm o}$, i.e.
\beq\label{ch5}
\p_{t_{2k_1}}\p_{t_{2k_2}}\ldots \p_{t_{2k_{2m+1}}}\log \tau^{\rm KP}\Bigr |_{{\bf t}_{\rm e}=0}=0
\eeq
for all $k_1, k_2, \ldots , k_{2m+1}\geq 1$, $m\geq 1$. Besides, first order derivatives
with respect to even times satisfy
\beq\label{ch6a}
\p_x\p_{t_4}\log \tau^{\rm KP}\Bigr |_{{\bf t}_{\rm e}=0}
=\p_x\p_{t_6}\log \tau^{\rm KP}\Bigr |_{{\bf t}_{\rm e}=0}
=\ldots =0.
\eeq
\end{lemma}

\noindent
The proof is given in Appendix A. From equations (\ref{ch6a}) we see that
$$
\p_{t_{2k}}\log \tau^{\rm KP}\Bigr |_{{\bf t}_{\rm e}=0}=\chi_{2k}(t_3, t_5, \ldots )
$$
does not depend on $x$. Equation (\ref{ch6}) means that
$\chi_2=0$. Next, from (\ref{ch5}) we conclude that
$$
\begin{array}{c}
\log \tau^{\rm KP}\Bigl (x, t_1, t_2,
t_3, t_4, \ldots \Bigr )
-\log \tau^{\rm KP}\Bigl (x, t_1, -t_2,
t_3, -t_4, \ldots \Bigr )=2\displaystyle{\sum_{k\geq 2}\chi_{2k}(t_3, t_5, \ldots )t_{2k}}
\end{array}
$$
is a linear function of ${\bf t}_{\rm e}$. Therefore, we can write
\beq\label{ch7}
\begin{array}{c}
\tau^{\rm KP}\Bigl (x, t_1, \frac{1}{2}\, z^{-2},
t_3, \frac{1}{4}\, z^{-4}, \ldots \Bigr )=
e^{2\chi_{\rm e}(t_3, t_5, \ldots ;z)}\tau^{\rm KP}\Bigl (x, t_1, -\frac{1}{2}\, z^{-2},
t_3, -\frac{1}{4}\, z^{-4}, \ldots \Bigr ),
\end{array}
\eeq
where $\chi_{\rm e}(t_3, t_5, \ldots ;z)$ is a function of the times $t_3, t_5, \ldots $ and
an even function of $z$. In its turn, (\ref{ch7}) implies
\beq\label{ch8}
\begin{array}{c}
\Psi^{\dag {\rm KP}}(x, {\bf \dot t}, z)=
C(t_3+\frac{1}{3}\, z^{-3}, t_5+\frac{1}{5}\, z^{-5}, \ldots ;z)
\Psi^{{\rm KP}}(x, {\bf \dot t}, -z),
\end{array}
\eeq
where $C(t_3, t_5, \ldots ;z)=e^{\chi_{\rm e}(t_3, t_5, \ldots ;z)}$ and
we use the short-hand notation $${\bf \dot t}=\{t_1, 0, t_3, 0, \ldots \}.$$
(In this notation equation (\ref{ch4}) takes the form
$\tau^{\rm KP}(x, {\bf \dot t}+[z^{-1}])=\tau^{\rm KP}(x, {\bf \dot t}-[-z^{-1}])$.)

The adjoint wave function $\Psi^{\dag {\rm KP}}$
satisfies the adjoint linear equation (see the independent proof
in the next section), which restricted to the locus ${\bf \dot t}$ where $B_k^{\dag}=-B_k$ for odd $k$ coincides with the linear equation for $\Psi^{{\rm KP}}$, so we simultaneously have
\beq\label{ch9}
\begin{array}{l}
\p_{t_k}\Psi^{\dag {\rm KP}}(x, {\bf \dot t}, z)=
B_k\Psi^{\dag {\rm KP}}(x, {\bf \dot t}, z),
\\ \\
\p_{t_k}\Psi^{{\rm KP}}(x, {\bf \dot t}, z)=
B_k\Psi^{{\rm KP}}(x, {\bf \dot t}, z).
\end{array}
\eeq
for odd $k$. Substituting (\ref{ch8}) into the first of these equations, we get,
after the change $z\to -z$,
$$
\begin{array}{c}
\p_{t_k}\Psi^{{\rm KP}}(x, {\bf \dot t}, z)
+\p_{t_k}\log C\Bigl (t_3-\frac{1}{3}\, z^{-3}, t_5-\frac{1}{5}\, z^{-5}, \ldots ;z\Bigr ) =B_k
\Psi^{{\rm KP}}(x, {\bf \dot t}, z),
\end{array}
$$
and from the second equation in (\ref{ch9}) we conclude that
$$
\begin{array}{c}
\p_{t_k}\log C\Bigl (t_3-\frac{1}{3}\, z^{-3}, t_5-\frac{1}{5}\, z^{-5}, \ldots ;z\Bigr )=0,
\end{array}
$$
i.e. $\chi_{\rm e}(t_3, t_5, \ldots ;z)=\chi_{\rm e}(z)$
is an even function of $z$ which does not depend on the times. (This function can be eliminated in
(\ref{ch7}) by passing to an equivalent tau-function.) Therefore, the equation (\ref{ch4})
which guarantees that $\tau^{\rm KP}$ is the KP extension of a solution to the CKP hierarchy
is proved.
%we can put $C(z)=1$,
%so the relation (\ref{ch4}) is established.

\medskip
\noindent
{\bf Remark.}
Passing to an equivalent tau-function using the transformation (\ref{linear}),
one obtains the condition $\p_{t_{2}}\log \tau^{\rm KP}\Bigr |_{{\bf t}_{\rm e}}=\gamma_{2}$
instead of (\ref{ch6}). Conversely, if
$\p_{t_{2}}\log \tau^{\rm KP}\Bigr |_{{\bf t}_{\rm e}}=\gamma_{2}$ with some nonzero
$\gamma_{2}$, it is possible to pass to an equivalent tau-function
satisfying (\ref{ch6}) by a transformation
of the form (\ref{linear}).

\medskip

In order to prove that $\tau =\sqrt{\vphantom{A^A}\tau^{\rm KP}}$  (\cite{CH14})
we compare two expressions for the wave
function $\Psi$ of the CKP hierarchy. The first one is
in terms of the KP tau-function (satisfying (\ref{ch6})),
\beq\label{e5a}
\Psi^{\rm KP}
= e^{xz+\zeta ({\bf t}_{\rm o}, z)}\frac{\tau^{\rm KP}(x, t_1-z^{-1}, -\frac{1}{2}\, z^{-2},
t_3-\frac{1}{3}\, z^{-3}, -\frac{1}{4}\, z^{-4}, \ldots )}{\tau^{\rm KP}
(t_1, 0, t_3, 0, \ldots )},
\eeq
and the second one (\ref{ckp6}) is in terms of the CKP tau-function $\tau$.
Recall that
\beq\label{e4a}
\Psi =z^{-1/2} \sqrt{\vphantom{B^{a^a}}\p_x \log \psi} \cdot \psi =
(2z)^{-1/2}\sqrt{\vphantom{B^{a^a}}\p_x \psi^2}
\eeq
(see (\ref{e4})),
where
\beq\label{e3a}
\psi = e^{xz+\zeta ({\bf t}_{\rm o}, z)}
\frac{\tau (x, {\bf t}_{\rm o}-2[z^{-1}]_{\rm o})}{\tau (x, {\bf t}_{\rm o})}.
\eeq
Comparing (\ref{e4a})
and (\ref{e5a}), we get the equation
\beq\label{e5b}
\frac{1}{2z}\, \p_x \!\! \left (e^{2xz}\frac{\tau^2 \Bigl (
x, {\bf t}_{\rm o}-2[z^{-1}]_{\rm o}\Bigr )}{\tau^2 (
x, {\bf t}_{\rm o})}\right )=
e^{2xz}\left (\frac{\tau^{\rm KP}(x, {\bf \dot t}-[z^{-1}])}{\tau^{\rm KP}
(x, {\bf \dot t})}\right )^2,
\eeq
where we again use the short-hand notation ${\bf \dot t}=\{t_1, 0, t_3, 0, \ldots \}$.
Then
using equation (\ref{ch3a}) we get that (\ref{e5b}) is equivalent to
the differential equation
\beq\label{e6}
\p_x \varphi = -2z \varphi ,
\eeq
where
\beq\label{e7}
\varphi = \frac{\tau^2 \Bigl (x, {\bf t}_{\rm o}-2[z^{-1}]_{\rm o}
\Bigr )}{\tau^2 (x, {\bf t}_{\rm o})}-
\frac{\tau^{\rm KP}\Bigl (x, {\bf \dot t}-2[z^{-1}]_{\rm o}\Bigr )}{\tau^{\rm KP}
(x, {\bf \dot t})}.
\eeq
In (\ref{e7}), ${\bf \dot t}-2[z^{-1}]_{\rm o}=\{t_1-2z^{-1}, 0, t_3-\frac{2}{3}\, z^{-3}, 0,
\ldots \}$.
The general solution of the differential equation (\ref{e6}) is
$$
\varphi =c(z, t_3, t_5, \ldots )e^{-2(x+t_1)z}
$$
but from (\ref{e7}) it follows that $\varphi$
is expanded in a power series
as $\varphi =\varphi_1 z^{-1}+\varphi_2 z^{-2}+\ldots$ as $z\to \infty$, and
this means that $c$ must be equal to 0. Therefore, $\varphi=0$, i.e.
\beq\label{e8}
\frac{\tau^2 \Bigl (x, {\bf t}_{\rm o}-2[z^{-1}]_{\rm o}
\Bigr )}{\tau^2 (x, {\bf t}_{\rm o})}=
\frac{\tau^{\rm KP}\Bigl (x, {\bf \dot t}-2[z^{-1}]_{\rm o}\Bigr )}{\tau^{\rm KP}
(x, {\bf \dot t})}
\eeq
for all $z$. This is an identity on solutions to the KP/CKP hierarchies. It follows from
(\ref{e8}) that $\tau^{\rm KP}={\rm const}\cdot \tau^2$, i.e.
$\tau (x, {\bf t}_{\rm o})=\sqrt{\tau^{\rm KP} (x, {\bf \dot t})}$ is a tau-function
of the CKP hierarchy.
\square

\medskip \noindent
{\bf Remark.} Equation (\ref{e5b}) is the CKP analog of the BKP statement that the corresponding
$\tau^{\rm KP}$ is a full square, i.e. $\tau = \sqrt{\tau^{\rm KP}\vphantom{A^A}}$
is an entire function of its variables. In the CKP case
$\p_x \psi^2$ is a full square.

\section{Algebraic-geometrical solutions to the KP and\\ CKP hierarchies}

The algebraic-geometrical construction of quasi-periodic solutions to
the CKP hierarchy briefly outlined in \cite{DJKM81} is a reduction
of the algebraic-geometrical construction of solutions to the KP hierarchy proposed in \cite{Krichever77,Krichever77a}. The main goal of this section is to give
a pure algebraic-geometrical proof of an identity for the Riemann theta-function
of a curve with involution having at least one fixed point. This identity is an algebraic-geometrical incarnation of the relations between KP and CKP tau-functions discussed in Section 2.

\subsection{Prelimineries}\label{sub:prel}

Let $\Gamma$ be a smooth compact algebraic curve of genus $g$. We fix a canonical basis of
cycles $a_{\alpha}, b_{\alpha}$ ($\alpha =1, \ldots , g$) with the intersections
$a_{\alpha}\circ a_{\beta}=b_{\alpha}\circ b_{\beta}=0$,
$a_{\alpha}\circ b_{\beta}=\delta_{\alpha \beta}$ and a basis of holomorphic
differentials $d\omega_{\alpha}$ normalized by the condition
$\displaystyle{\oint_{a_{\alpha}}d\omega_{\beta}=\delta_{\alpha \beta}}$.
The period matrix is defined as
\beq\label{qp1}
T_{\alpha \beta}=\oint_{b_{\alpha}}d\omega_{\beta}, \qquad \alpha , \beta =1, \ldots , g.
\eeq
It is a symmetric matrix with positively defined imaginary part.
The Riemann theta-function is defined by the series
\beq\label{qp2}
\theta(\vec z)=\theta(\vec z|T)=
\sum_{\vec n \in \z ^{g}}e^{\pi i (\vec n, T\vec n)+2\pi i (\vec n, \vec z)},
\eeq
where $\vec z=(z_1, \ldots , z_g)$ and $\displaystyle{(\vec n, \vec z)=
\sum_{\alpha =1}^g n_{\alpha}z_{\alpha}}$.

The Jacobian of the curve $\Gamma$ is the $g$-dimensional complex torus
\beq\label{qp4}
J(\Gamma )=\CC ^g /\{2\pi i \vec N +2\pi i T \vec M\},
\eeq
where $\vec N$, $\vec M$ are $g$-dimensional vectors with integer components.
Fix a point $Q_0\in \Gamma$ and define the Abel map $\vec A(P)$, $P\in \Gamma$
from $\Gamma$ to $J(\Gamma )$, as
\beq\label{qp3}
\vec A(P)=\vec \omega (P)=
\int_{Q_0}^P d \vec \omega , \qquad d\vec \omega =(d\omega_1, \ldots , d\omega_g ).
\eeq
The Abel map can be extended to the group of divisors ${\cal D}=n_1Q_1+\ldots +n_KQ_K$ as
\beq\label{qp3a}
\vec A({\cal D})=\sum_{i=1}^K n_i\int_{Q_0}^{Q_i} d \vec \omega =\sum_{i=1}^K
n_i\vec A(Q_i).
\eeq

Let $P_{\infty}\in \Gamma$ be a marked point and $k^{-1}$ a local parameter
in a neighborhood of the marked point ($k=\infty$ at $P_{\infty}$). Let $d\Omega_j$
be abelian differentials of the second kind with the only pole at $P_{\infty}$ of the form
\beq\label{defdomega}
d\Omega_j = dk^j +O(k^{-2})dk, \quad k\to \infty
\eeq
normalized by the condition $\displaystyle{\oint_{a_{\alpha}}d\Omega_{j}=0}$, and $\Omega_j$
be the (multi-valued) functions
$$
\Omega_j(P)=\int_{Q_0}^{P}d\Omega_j +q_j,
$$
where the constants $q_j$ are chosen in such a way that $\Omega_i(P)=k^i +O(k^{-1})$, namely,
\beq\label{qp3b}
\Omega_i(P)=k^i +\sum_{j\geq 1} \frac{1}{j}\, \Omega_{ij}k^{-j}.
\eeq
The Riemann identity implies that the matrix $\Omega_{ij}$ is symmetric:
$\Omega_{ij}=\Omega_{ji}$.

Set
\beq\label{qp5}
U_j^{\alpha}=\frac{1}{2\pi i}\oint_{b_{\alpha}}d\Omega_j, \qquad
\vec U_j =(U_j^{1}, \ldots , U_j^g).
\eeq
One can prove the following relation:
\beq\label{qp6a}
\vec d\omega =\sum_{j\geq 1}\vec U_j k^{-j-1}dk
\eeq
or
\beq\label{qp6}
\vec A(P)-\vec A(P_{\infty})=\int_{P_{\infty}}^P d\vec \omega =
-\sum_{j\geq 1}\frac{1}{j}\, \vec U_j k^{-j}.
\eeq

We will also use the following fact \cite{Fay,Mumford}: for any non-special effective divisor
${\cal D}=Q_1+\ldots +Q_g$ of degree $g$ the function
$$
f(P)=\theta \Bigl (\vec A(P)-\vec A({\cal D}) -\vec K\Bigr )
$$
has exactly $g$ zeros at the points $Q_1, \ldots , Q_g$. Here $\vec K=
(K_1, \ldots , K_g)$ is the
vector of Riemann's constants
\beq\label{qp7}
K_{\alpha}= \pi i +\pi i T_{\alpha \alpha}-2\pi i
\sum_{\beta \neq \alpha}\oint_{a_{\beta}}\omega_{\alpha} (P)d\omega_{\beta}(P).
\eeq
Let ${\cal K}$ be the canonical class of divisors (the equivalence class of divisors
of poles and zeros of abelian differentials on $\Gamma$), then one can show that
\beq\label{qp8}
2\vec K=-\vec A({\cal K}).
\eeq
It is known that $\mbox{deg}\, {\cal K}=2g-2$. In particular, this means that
holomorphic differentials have $2g-2$ zeros on $\Gamma$.

We also need the bi-differential $d_P d_Q \Omega (P, Q)$ such that it is symmetric in
$P, Q$, its
only singularity is a second order
pole at $P=Q$ and the integrals over $a$-cycles vanish. It is related to the
differentials $d\Omega _j$ as follows:
\beq\label{qp81}
\res_{Q=P_{\infty}} \Bigl (k^i(Q)d_P d_Q \Omega (P, Q)\Bigr ) =-d\Omega_i (P).
\eeq
The expansion in the local parameters is
\beq\label{qp82}
d_P d_Q \Omega (P, Q)=\Biggl ( \frac{1}{(k^{-1}(P)\! -\! k^{-1}(Q))^2}-
\sum_{i,j\geq 1} \Omega_{ij} k^{1-i}(P)k^{1-j}(Q)\Biggr ) dk^{-1}(P)dk^{-1}(Q).
\eeq
In fact this bi-differential can be expressed in terms of the odd theta-function
$$\theta_{*}(\vec z)=\theta \left [\begin{array}{c}\vec \delta ' \\ \vec \delta^{''}
\end{array}\right ] (\vec z)
=\sum_{\vec n \in \z ^{g}}e^{\pi i (\vec n +\vec \delta ',
T(\vec n +\vec \delta '))+2\pi i (\vec n+\vec \delta ', \vec z+\vec \delta '')},$$
where $(\vec \delta ', \vec \delta '')$ is a non-singular odd theta-characteristics.
One has:
\beq\label{qp83}
d_P d_Q \Omega (P, Q)=d_P d_Q \log \theta_{*}\Bigl (\vec A(P)-\vec A(Q)\Bigr ).
\eeq
Calculating the double integral
$$
\int_{P_{\infty}}^{P_1}\int_{Q_0}^{P_2}d_P d_Q \Omega (P, Q)
$$
in two ways (using first (\ref{qp82}) and then (\ref{qp83})), we obtain the equality
$$
\log \frac{(k^{-1}(P_1)-k^{-1}(P_2))
k^{-1}(Q_0)}{(k^{-1}(P_1)-k^{-1}(Q_0))k^{-1}(P_2)}
-\sum_{i,j\geq 1} \Omega_{ij} \frac{k^{-i}(P_1)k^{-j}(P_2)}{ij}+
\sum_{i,j\geq 1} \Omega_{ij} \frac{k^{-i}(P_1)k^{-j}(Q_0)}{ij}
$$
$$
=\log \frac{\theta_{*}\Bigl (\vec A(P_2)-\vec A(P_1)\Bigr )
\theta_{*}\Bigl (\vec A(P_\infty )\Bigr )}{\theta_{*}\Bigl (\vec A(P_2)-
\vec A(P_\infty )\Bigr )\theta_{*}\Bigl (\vec A(P_1 )\Bigr )}
$$
Tending here $Q_0\to P_{\infty}$, we arrive at the important relation
\beq\label{qp84}
\exp \Biggl (-\sum_{i,j\geq 1}\Omega_{ij}\frac{k_1^{-i}k_2^{-j}}{ij}\Biggr )=
\frac{C \theta_{*}\Bigl ( \vec A(P_1)-\vec A(P_2)\Bigr )}{(k_1-k_2)
\theta_{*}\Bigl ( \vec A(P_1)\! -\! \vec A(P_\infty )\Bigr )
\theta_{*}\Bigl ( \vec A(P_2)\! -\! \vec A(P_\infty )\Bigr )},
\eeq
where
$$C=\sum_{\alpha =1}^{g}U_1^{\alpha} \theta_{*, \alpha} (\vec 0), \qquad
\theta_{*, \alpha}(\vec 0)=\frac{\p \theta_{*}(\vec z)}{\p z_{\alpha}}\Biggr |_{\vec z=0}
$$
is a constant and $k_1=k(P_1), k_2=k(P_2)$. In particular, tending
$k_1\to k_2$, we get
\beq\label{qp85}
\exp \Biggl (-\sum_{i,j\geq 1}\Omega_{ij}\frac{k^{-i-j}}{ij}\Biggr )dk=
\frac{Cd\zeta}{\theta_{*}^2\Bigl ( \vec A(P)\! -\! \vec A(P_\infty )\Bigr )},
\eeq
where $d\zeta$ is the holomorphic differential
\beq\label{qp86}
d\zeta =\sum_{\alpha =1}^g \theta_{*, \alpha}(\vec 0)d\omega_{\alpha}.
\eeq
As is explained in \cite{Mumford}, the differential $d\zeta$ has double zeros at
$g-1$ points $R_1, \ldots , R_{g-1}$ while the function
$$
f_{*}(P)= \theta_{*}\Bigl ( \vec A(P)\! -\! \vec A(P_\infty )\Bigr )
$$
has simple zeros at the same points $R_i$ and $P_{\infty}$. Therefore, the differential
in the right hand side of (\ref{qp85}) has the only (second order) pole at $P_{\infty}$
and no zeros. However, this differential is well-defined only on a covering of the curve
$\Gamma$ because it is not single-valued.

Finally, we mention the trisecant Fay identity \cite{Fay}:
\beq\label{Fay}
\begin{array}{c}
\theta_{*}\Bigl (\vec A(P_1)-\vec A(P_2)\Bigr )
\theta_{*}\Bigl (\vec A(P_3)-\vec A(P_4)\Bigr )
\theta \Bigl (\vec z +\vec A(P_1)+\vec A(P_2)\Bigr )
\theta \Bigl (\vec z +\vec A(P_3)+\vec A(P_4)\Bigr )
\\ \\
+\theta_{*}\Bigl (\vec A(P_2)-\vec A(P_3)\Bigr )
\theta_{*}\Bigl (\vec A(P_1)-\vec A(P_4)\Bigr )
\theta \Bigl (\vec z +\vec A(P_2)+\vec A(P_3)\Bigr )
\theta \Bigl (\vec z +\vec A(P_1)+\vec A(P_4)\Bigr )
\\ \\
+ \theta_{*}\Bigl (\vec A(P_3)\! -\!\vec A(P_1)\Bigr )
\theta_{*}\Bigl (\vec A(P_2)\! -\! \vec A(P_4)\Bigr )
\theta \Bigl (\vec z \! +\! \vec A(P_1)\! +\! \vec A(P_3)\Bigr )
\theta \Bigl (\vec z \! +\! \vec A(P_2)\! +\! \vec A(P_4)\Bigr )=0.
\end{array}
\eeq

\subsection{The Baker-Akhiezer function}

Let $x, t_1, t_2, t_3, \ldots$ be a set of complex parameters
(here we assume that only a finite
number of them are different from zero) and let $\Gamma$ be a smooth
genus $g$ algebraic curve with fixed local coordinate $k^{-1}(P)$ in the
neighborhood of a fixed point $P_\infty$.

\begin{lemma}\label{lm:BA} (\cite{Krichever77,Krichever77a}) Let
${\cal D}=Q_1+ \ldots + Q_g$ be an effective non-special divisor of
degree $g$. Then there is the unique function $\Psi_{BA}(x,{\bf t},P)$ such that:

$1^0$. As a function of $P\in \Gamma$ it is meromorphic away from the marked
point $P_\infty$ with poles at the points $Q_s$ of multiplicity not
greater then the multiplicity of $Q_s$ in $\cal D$.

$2^0$. In the neighborhood  of $P_{\infty}$ it has the form

\beq\label{ba101}
\Psi_{BA} = \exp \Bigl (xk+\sum_{j\geq 1}t_j k^j\Bigr )
\Bigl (1+\xi_1 k^{-1} +\xi_2 k^{-2} +\ldots \Bigr ), \ \ \ k=k(P).
\eeq
\end{lemma}
The function $\Psi_{BA}$ is called (one-point) Baker-Akhiezer function.

An easy corollary of the uniqueness of the Baker-Akhiezer function is
\begin{theorem}\label{thm:BA}(\cite{Krichever77,Krichever77a})
Let $\Psi_{BA}$ be the Baker-Akhiezer function defined by
Lemma \ref{lm:BA}. Then for each $j=1,2,3, \ldots$ there is a unique
differential operator $B_j$ such that the equation
\beq\label{w5}
(\p_{t_j}-B_j)\Psi_{BA} =0
\eeq
holds.
\end{theorem}
The operators $B_j$ above can be easy expressed in terms of the dressing operator $W$ for the Baker-Akhiezer function. Namely, the infinite series (\ref{ba1}) can be represented as
\beq\label{w1}
\Psi_{BA} = W \exp \Bigl (xk+\sum_{j\geq 1}t_j k^j\Bigr ),
\eeq
where $W$ is of the form (\ref{ckp1a}). The corresponding Lax operator of the KP
hierarchy is $\LL=W\p_xW^{-1}$.
By the definition we have
\beq\label{w4}
\LL\Psi_{BA} =k\Psi_{BA} .
\eeq
The operator $B_j$ in Theorem \ref{thm:BA} was defined as the unique
monic order $j$ operator such that the congruence
$$ (k^j-B_j) \Psi_{BA}=O(1/k)\exp \Bigl (xk+\sum_{j\geq 1}t_j k^j\Bigr )$$
holds. Using (\ref{w4}), it is easy to identify $B_j=\LL^j_+$. Indeed
$$(k^j-\LL^j_+)\Psi_{BA} = (\LL^j-\LL_+^j)\Psi_{BA} =\LL_-^j\Psi_{BA}=O(1/k)\exp \Bigl (xk+\sum_{j\geq 1}t_j k^j\Bigr )
$$
The compatibility conditions of equations (\ref{w5}) imply
\begin{corollary} The operators $B_j$ defined by the BA function satisfies the equations
\beq\label{w6}
[\p_{t_j}-B_j, \p_{t_l}-B_l]=0.
\eeq
\end{corollary}
It is the Zakharov-Shabat form (\ref{kp10}) of the KP hierarchy.
Note that equation (\ref{w5}) implies the evolution equation for the
dressing operator:
\beq\label{w6a}
\p_{t_j}W=-(W\p_x^j W^{-1})_{-}W,
\eeq
where $(\ldots )_{-}$ is the projection to negative powers of the operator $\p_x$.

\subsection{The dual Baker-Akhiezer function}

For further comparison with the tau-functional formulation
of the KP hierarchy let us present the notion of the {\it dual} (adjoint)
Baker-Akhiezer function introduced in \cite{Cherednik} (see the 
details in \cite{kp1,handbook}).

First we define duality for divisors of degree $g$.
For a generic effective degree $g$ divisor ${\cal D}=Q_1+\ldots+Q_g$ there
is a unique up to a constant factor the abelian differential $d\Omega$ with
the only (second order) pole at $P_{\infty}$ vanishing (with the
corresponding multiplicity) at the points $Q_s$. The zero
divisor of $d\Omega$ is of degree $2g$. Hence, it has other $g$ zeros at some points
$Q_1^{\dag}, \ldots , Q_g^{\dag}$. The divisor
$${\cal D}^{\dag}=Q_1^{\dag}+ \ldots + Q_g^{\dag}$$
is called {\it dual to $\cal D$}.
By the definition we have the equality
\beq\label{ba201}
{\cal D}+{\cal D}^{\dag}={\cal K}+2P_{\infty}
\eeq
(where ${\cal K}$ is the canonical class),
which under the Abel transform takes the form
\beq\label{ba301}
\vec A({\cal D})+\vec A({\cal D}^{\dag})+2\vec K -2\vec A(P_{\infty})=0
\eeq

The dual (adjoint) Baker-Akhiezer function
$\Psi^{\dag}_{BA}$ has the divisor of poles ${\cal D}^{\dag}$ and
in the vicinity of $P_{\infty}$ it
has the form
\beq\label{ba401}
\Psi^{\dag}_{BA} = \exp \Bigl (-xk-\sum_{j\geq 1}t_j k^j\Bigr )
\Bigl (1+\xi_1^{\dag} k^{-1} +\xi_2^{\dag} k^{-2} +\ldots \Bigr ).
\eeq

The differential $\Psi_{BA} (x,{\bf t}, P)\Psi_{BA}^{\dag}(x, {\bf t}', P)d\Omega (P)$,
where we have denoted the set of times
as ${\bf t}=\{ t_1, t_2, t_3, \ldots \}$ for brevity, is holomorphic
everywhere on $\Gamma$ except the point $P_{\infty}$ (because poles of the Baker-Akhiezer
functions are canceled by zeros of $d\Omega$). Therefore,  its ``residue'' at this point
is equal to zero, i.e.,
\beq\label{ba701}
\oint_{C_{\infty}} \! \Psi_{BA} (x,{\bf t}, P)\Psi_{BA}^{\dag}(x, {\bf t}', P)d\Omega (P) =0
\eeq
for all ${\bf t}, {\bf t}'$, where $C_{\infty}$ is a small contour around the point $P_{\infty}$.
Equation (\ref{ba701}) is equivalent to the equation
\beq\label{ba701b}
\res_{P_\infty}\,\left(\p_x^i \Psi_{BA} (x,{\bf t}, P)
\Psi_{BA}^{\dag}(x, {\bf t}, P)\right)d\Omega (P)=0,\,\ \ i=1,2,3,\ldots
\eeq
from which one can derive the following theorem (see \cite{kp1}
and \cite{handbook} for more details).

\begin{theorem}
The dual Baker-Akhiezer function is equal to
\beq\label{ad1}
\Psi_{BA}^{\dag}=(W^{\dag})^{-1} \exp \Bigl (-xk -\sum_{j\geq 1}t_jk^j\Bigr )
\eeq
and satisfies the adjoint equations
\beq\label{ad2}
\LL^{\dag}\Psi_{BA}^{\dag}=k\Psi_{BA}^{\dag},
\quad -\p_{t_j}\Psi_{BA}^{\dag}=B_j^{\dag}\Psi_{BA}^{\dag}.
\eeq
\end{theorem}

\noindent
For completeness we outline here a direct proof of the theorem.
Equations (\ref{ad2}) immediately follow from (\ref{w6a}) and (\ref{ad1}):
$$
-\p_{t_j}\Psi_{BA}^{\dag}=\Bigl (k^j (W^{\dag})^{-1}-(W^{\dag})^{-1}\p_{t_j}W^{\dag}(W^{\dag})^{-1}
\Bigr )\exp \Bigl (-xk -\sum_{j\geq 1}t_jk^j\Bigr )
$$
$$
=\Bigl (k^j (W^{\dag})^{-1}-((W^{\dag})^{-1}(-\p_x )^j W^{\dag})_- (W^{\dag})^{-1}
\Bigr )\exp \Bigl (-xk -\sum_{j\geq 1}t_jk^j\Bigr )
$$
$$
=\Bigl ((\LL^{\dag})^j -(\LL^{\dag})^j_-\Bigr ) \Psi_{BA}^{\dag}=B_j^{\dag}\Psi_{BA}^{\dag}.
$$
In order to prove (\ref{ad1}) we note that equation (\ref{ba701}) written in the local
parameter $k$ implies
$$
b_m=\frac{1}{2\pi i}\,
\p_{x'}^m \oint_{C_{\infty}}\! \Psi_{BA} (x, {\bf t}, k)\Psi_{BA}^{\dag} (x', {\bf t}, k)
\varphi (k)\frac{dk}{2\pi i} \Biggr |_{x'=x}=0 \quad \mbox{for all $m\geq 0$.}
$$
Here
$$
\varphi (k) = \frac{d\Omega}{dk} =\sum_{j\geq 0}\varphi_j k^{-j}.
$$
We set
$$
\Psi_{BA}^{\dag}=V\exp \Bigl (-xk -\sum_{j\geq 1}t_jk^j\Bigr ), \quad
V=1+\xi_1^{\dag}\p_x^{-1}+\xi_2^{\dag}\p_x^{-2}+\ldots
$$
We have:
$$
b_m=\frac{1}{2\pi i}\oint_{C_{\infty}}
\Bigl (\sum_{l\geq 0}\varphi_l k^{-l}\Bigr )
\Bigl (\sum_{j\geq 0}\xi_j(x)k^{-j}\Bigr )\p_{x'}^m
\Bigl (\sum_{i\geq 0}\xi_i^{\dag}(x')(-k)^{-i}\Bigr )e^{(x-x')z} \frac{dk}{2\pi i} \Biggr |_{x'=x}
$$
$$
=\frac{1}{2\pi i}\oint_{C_{\infty}}
\Bigl (\sum_{l\geq 0}\varphi_l k^{-l}\Bigr )
\Bigl (\sum_{j\geq 0}\xi_j k^{-j}\Bigr )(\p_x -k)^m
\Bigl (\sum_{i\geq 0}\xi_i (-k)^{-i}\Bigr )\frac{dk}{2\pi i}
$$
$$
=\sum_{l=0}^m (-1)^{l}\frac{m!}{(m-l)!}\, \varphi_l
\! \sum_{i+j+s=m-l+1}(-1)^{m-l+i+s}\left (\! \begin{array}{c}m-l\\ s \end{array} \! \right )
\xi_j \p_x^s \xi^{\dag}_i.
$$
But the last sum is the coefficient of $(-1)^m \p_x^{-m+l-1}$ in the operator
$WV^{\dag}$, so we can write:
$$
b_m= \sum_{l=0}^m (-1)^{m-l}\frac{m!}{(m-l)!}\, \varphi_l
\Bigl (W V^{\dag}\Bigr )_{-m+l-1}=0 \quad \mbox{for all $m\geq 0$.}
$$
This is a homogeneous triangular system of linear equations for the coefficients
$\Bigl (W V^{\dag}\Bigr )_{-l}$. The unique solution is
$\Bigl (W V^{\dag}\Bigr )_{-l}=0$ for all $l\geq 1$, hence
$WV^{\dag}=1$, i.e. $V=(W^{\dag})^{-1}$.
\square

\subsection{Theta-functional formulae}

The Baker-Akhiezer function can be explicitly written in terms of the Riemann-theta function \cite{Krichever77}:
\beq\label{ba501}
\begin{array}{l}
\displaystyle{
\Psi_{BA} (P)= \exp \Bigl (x\Omega_1(P)+\! \sum_{j\geq 1}t_j \Omega_j (P)\Bigr )}
\\ \\
\displaystyle{
\phantom{aaaaaaaaaa}\times \,
\frac{\theta \Bigl (\vec A(P)+\vec U_1x+\sum\limits_{j\geq 1}\vec U_j t_j -
\vec A({\cal D}) -\vec K\Bigr )\theta \Bigl (\vec A({\cal D}) +
\vec K-\vec A(P_{\infty})\Bigr )}{\theta \Bigl (\vec A(P)-
\vec A({\cal D}) -\vec K\Bigr )\theta \Bigl (\vec U_1x+\sum\limits_{j\geq 1}\vec U_j t_j
-\vec A({\cal D}) -
\vec K+\vec A(P_{\infty})\Bigr ) }}.
\end{array}
\eeq
For the proof of (\ref{ba501}) it is enough to check that the right-hand
side of (\ref{ba501}) $(a)$ is a single-valued function on $\Gamma$, i.e.
does not depend on the choice of path of integration in the definition of
the Abel map and the Abelian integral $\Omega_j$  (which is assumed to be
the same for the both objects); $(b)$ has required exponential singularity
at the marked point $P_\infty$; $(c)$ outside of $P_\infty$ is meromorphic
with divisor of poles at the divisor $\cal D$.

The proof of $(a)$ directly follows from the monodromy properties of the
theta-function and the definition (\ref{qp5}) of the vectors $U_j$. The
proof of $(b)$ follows from the definition of the differentials $d\Omega_j$.
The $(c)$ part follows from the Jacoby inversion theorem above.

The corresponding expression for the adjoint Baker-Akhiezer function is
\beq\label{ba5a01}
\begin{array}{l}
\displaystyle{
\Psi_{BA}^{\dag} (P)= \exp \Bigl (-x\Omega_1(P)-\! \sum_{j\geq 1}t_j \Omega_j (P)\Bigr )}
\\ \\
\displaystyle{
\phantom{aaaaaaaaaa}\times \,
\frac{\theta \Bigl (\vec A(P)-\vec U_1x-\sum\limits_{j\geq 1}\vec U_j t_j -
\vec A({\cal D}^{\dag}) -\vec K\Bigr )\theta \Bigl (\vec A({\cal D}^{\dag}) +
\vec K-\vec A(P_{\infty})\Bigr )}{\theta \Bigl (\vec A(P)-
\vec A({\cal D}^{\dag}) -\vec K\Bigr )\theta \Bigl (\vec U_1x+\sum\limits_{j\geq 1}\vec U_j t_j
+\vec A({\cal D}^{\dag}) +
\vec K-\vec A(P_{\infty})\Bigr ) }}.
\end{array}
\eeq
Using the relation (\ref{ba3}), one can rewrite it in the form
\beq\label{ba601}
\begin{array}{l}
\displaystyle{
\Psi_{BA}^{\dag} (P)= \exp \Bigl (-x\Omega_1(P)-\! \sum_{j\geq 1}t_j \Omega_j (P)\Bigr )}
\\ \\
\displaystyle{
\phantom{aaaaa}\times \,
\frac{\theta \Bigl (\vec A(P)-\vec U_1x-\sum\limits_{j\geq 1}\vec U_j t_j +
\vec A({\cal D}) +\vec K -2\vec A(P_{\infty})\Bigr )\theta \Bigl (\vec A({\cal D}) +
\vec K-\vec A(P_{\infty})\Bigr )}{\theta \Bigl (\vec A(P)+
\vec A({\cal D}) +\vec K-2\vec A(P_{\infty})\Bigr )
\theta \Bigl (\vec U_1x+\sum\limits_{j\geq 1}\vec U_j t_j
-\vec A({\cal D}) -
\vec K+\vec A(P_{\infty})\Bigr ) }}.
\end{array}
\eeq

In order to find the solution $u_1=u_1(x, t_1, t_2, t_3, \ldots )$ to the KP hierarchy,
one should find the coefficient $\xi_1$ in the expansion
$$
\log \Psi_{BA} = xk +\sum_{j\geq 1}t_jk^j +\xi_1 k^{-1}+O(k^{-2}).
$$
A direct calculation with the help of the explicit formula (\ref{ba501}) yields
\beq\label{xi1}
\xi_1 = x\Omega_{11} +\sum_{i\geq 1}t_i \Omega_{1i} -\p_x \log \theta \Bigl (
\vec U_1 x +\sum_{j\geq 1}\vec U_j t_j +\vec Z\Bigr ) + \mbox{const},
\eeq
where $\vec Z = -\vec A({\cal D})-\vec K +\vec A(P_{\infty})$. Therefore,
\beq\label{xi2}
u_1=-\xi_1'= \p_x^2 \log \theta \Bigl (
\vec U_1 x +\sum_{j\geq 1}\vec U_j t_j +\vec Z\Bigr )-\Omega_{11}.
\eeq

\subsection{The tau-function}

Without loss of generality we can put $x=0$ for simplicity.
The dependence on $x$ can be restored by the substitution $t_1\to t_1+x$.

The theta-functional formula (\ref{ba501}) for the Baker-Akhiezer function
and the expansion (\ref{qp6}) of the Abel map near $P_\infty$ allows to
reformulate the above presented construction of algebraic-geometrical
construction in terms of the tau-functional formulation of the KP hierarchy. Namely,
we have the following theorem.

\begin{theorem} (\cite{Krichever77,Dubrovin81})
The right hand side of the equation
\beq\label{tau1}
\tau^{\rm KP} ({\bf t})=\exp \Bigl ( -
\frac{1}{2}\sum_{i,j\geq 1}\Omega_{ij}t_it_j\Bigr )\,
\theta \Bigl (\sum_{j\geq 1}\vec U_j t_j +\vec Z \Bigr ),
\eeq
where the constant vector $\vec Z$ is parameterized through the divisor ${\cal D}$ as
\beq\label{tau1a}
\vec Z=-\vec A({\cal D})-\vec K
+\vec A(P_{\infty})
\eeq
is the KP tau-function.
\end{theorem}

\noindent
{\it Proof.}
Equation (\ref{qp6}) implies that
$$
\begin{array}{c}
\displaystyle{
\theta \Bigl (\sum_{j\geq 1}\vec U_j (t_j \mp \frac{1}{j}\, k^{-j}) +\vec Z \Bigr )
=\theta \Bigl (\pm \vec A(P)+\sum_{j\geq 1}\vec U_j t_j  +\vec Z \mp \vec A(P_{\infty}) \Bigr ),}
\end{array}
$$
so we see that the Baker-Akhiezer functions (\ref{ba501}), (\ref{ba601})
are connected with the tau-function by the standard formulas \cite{JimboMiwa,DJKM83}
\beq\label{tau2}
\Psi_{BA} = C(k)\exp \Bigl (\sum_{j\geq 1}t_jk^j\Bigr )
\frac{\tau^{\rm KP} ({\bf t}-[k^{-1}] )}{\tau^{\rm KP} ({\bf t} )},
\eeq
\beq\label{tau3}
\Psi_{BA}^{\dag} = C^{\dag}(k)\exp \Bigl (-\sum_{j\geq 1}t_jk^j\Bigr )
\frac{\tau^{\rm KP} ({\bf t}+[k^{-1}] )}{\tau^{\rm KP} ({\bf t} )},
\eeq
where $C(k)$, $C^{\dag}(k)$ are normalization factors such that
$C(k)=1+O(k^{-1})$, $C^{\dag}(k)=1+O(k^{-1})$.
A simple calculation shows that
\beq\label{tau6a}
\begin{array}{l}
\displaystyle{
\frac{\tau^{\rm KP} ({\bf t}-[k_1^{-1}]-[k_2^{-1}])\tau^{\rm KP}
({\bf t})}{\tau^{\rm KP} ({\bf t}-[k_1^{-1}])
\tau^{\rm KP} ({\bf t}-[k_2^{-1}])}=
\exp \Biggl (-\! \sum_{i,j\geq 1}\Omega_{ij}\frac{k_1^{-i}k_2^{-j}}{ij}\Biggr )}
\\ \\
\phantom{aaaaaaaaaaaaaa}\displaystyle{\times \,
\frac{\theta \Bigl (\vec A(P_1)+\vec A(P_2) +\sum\limits_{j\geq 1}\vec U_j t_j +
\vec Z \Bigr )\theta \Bigl (\sum\limits_{j\geq 1}\vec U_j t_j +
\vec Z \Bigr )}{\theta \Bigl (\vec A(P_1)+\sum\limits_{j\geq 1}\vec U_j t_j +
\vec Z \Bigr )\theta \Bigl (\vec A(P_2) +\sum\limits_{j\geq 1}\vec U_j t_j +
\vec Z \Bigr )}}.
\end{array}
\eeq
Using (\ref{qp84}), it is straightforward to check that the tau-function
(\ref{tau1}) satisfies the Hirota-Miwa equation (\ref{tau6b})
which is the generating equation for the KP hierarchy. It appears to be equivalent to the
Fay identity (\ref{Fay}).
\square

It is interesting to compare equation (\ref{ba701}) and the bilinear
relation (\ref{ch02}) for the tau-function. They coincide if
\beq\label{tau5}
\begin{array}{c}
\displaystyle{
d\Omega =\frac{\tau^{\rm KP} (-[k^{-1}])\tau^{\rm KP} ([k^{-1}])}{(\tau^{\rm KP}(0))^2}}\, dk
\\ \\
\displaystyle{
=
\frac{\theta \Bigl (\vec A (P) -\vec A({\cal D}) -\vec K\Bigr )
\theta \Bigl (\vec A (P) -\vec A({\cal D}^{\dag}) -
\vec K)\Bigr )}{\theta^2 \Bigl (\vec A({\cal D}) +
\vec K-\vec A(P_{\infty})\Bigr )}
\exp \Bigl (-\! \sum_{i,j\geq 1}\Omega_{ij}\, \frac{k^{-i-j}}{ij}\Bigr ) \, dk.}
\end{array}
\eeq
Using (\ref{qp85}), we can rewrite this as
\beq\label{tau7}
d\Omega =
C\frac{\theta \Bigl (\vec A (P) -\vec A({\cal D}) -\vec K\Bigr )
\theta \Bigl (\vec A (P) -\vec A({\cal D}^{\dag}) -
\vec K)\Bigr )}{\theta_{*}^2\Bigl ( \vec A(P)\! -\! \vec A(P_\infty )\Bigr )} \, d\zeta ,
\eeq
where the holomorphic differential $d\zeta$ is given by (\ref{qp86}). Its properties
(see \cite{Mumford}) imply that the differential
in the right hand side is a well-defined meromorphic differential on $\Gamma$ with the
only second order pole at $P_{\infty}$ and $2g$ zeros at the points of the divisors
${\cal D}$, ${\cal D}^{\dag}$. Therefore, it has all the properties of the
differential $d\Omega$ and hence must be proportional to it. The equality (\ref{tau7})
just reflects this fact.

\medskip \noindent
{\bf Remark.}
The function $\Psi_{BA}$ and the wave function $\Psi$ introduced in section \ref{section:wave}
(see (\ref{ckp4a})) differ by a normalization factor depending on $k(P)$. From (\ref{tau5})
it follows that
\beq\label{tau7101}
\Psi_{BA}(x,{\bf t}, P)\Psi^{\dag}_{BA}(x,{\bf t}', P)d\Omega =
\Psi (x,{\bf t}, k)\Psi^{\dag}(x,{\bf t}', k)dk.
\eeq

\subsection{Curves with involution: solutions to the CKP hierarchy}

Let $\Gamma$ be a smooth genus $g$ algebraic curve with involution $\iota$
having $2(n+1)>0$ fixed points. By the Riemann-Hurwitz formula $g=2g_0+n$
where $g_0$ is the genus of the factor-curve $\Gamma_0=\Gamma/\iota$. It is
known that on  $\G$ there is a basis of $a$- and $b$-cycles  with canonical
intersection matrix: $a_i\cdot a_j=b_i\cdot b_j=0, a_i\cdot b_j=\delta_{ij};$ and such that
in this basis the action of the involution $\iota$ has the form
\beq\label{sa}\iota(a_i)=a_{i+g_0}, \ \ \iota(b_i)=b_{i+g_0}, \ i=1,\ldots, g_0,
\eeq
and
\beq\label{sa1}\iota(a_i)=-a_i, \ \ \iota(b_i)=-b_i, \ i=2g_0+1, \ldots, 2g_0+n.
\eeq
Let the marked point $P_\infty$ on $\Gamma$ be one of the fixed points of the involution, $\iota(P_\infty)=P_\infty$ and let $z=k^{-1}$ be a local coordinate in the neighborhood
of $P_\infty$ that is odd with respect to the involution, $\iota^* (k)=-k$.
From the definition of the abelian differentials
$d\Omega _j$ in subsection \ref{sub:prel} it follows that
\beq\label{iota0}
d\Omega_j (\iota P)=(-1)^jd\Omega_j (P)
\eeq
and, therefore,
\beq\label{iota4}
\Omega_j (\iota P)=(-1)^j \, \Omega_j(P).
\eeq
Suppose that the divisor ${\cal D}$ satisfies the constraint
\beq\label{iota1}
{\cal D}+\iota {\cal D}={\cal K}+2P_{\infty}.
\eeq
Then for the Baker-Akhiezer function defined by $\Gamma, P_\infty$,
the local coordinate $k^{-1}$ and the divisor $\cal D$ the equation
\beq\label{iota2}
\Psi_{BA}^{\dag} (t_1, 0, t_3, 0, \ldots , P)=\Psi_{BA} (t_1, 0, t_3, 0, \ldots , \iota P).
\eeq
holds. The bilinear relation (\ref{ba701}) takes the form
\beq\label{ba701a}
\oint_{C_{\infty}} \! \Psi_{BA} (x,{\bf t}_{\rm o}, P)
\Psi_{BA} (x, {\bf t}_{\rm o}', \iota P)d\Omega (P) =0
\eeq
for all ${\bf t}_{\rm o}, {\bf t}_{\rm o}'$.

Using formulas (\ref{ba501}), (\ref{ba601}) we can write the relation
(\ref{iota2}) in the explicit form:
\beq\label{iota3}
\begin{array}{l}
\displaystyle{
\Psi_{BA} (t_1, 0, t_3, 0, \ldots , \iota P)=
\exp \Bigl (\sum_{j\geq 1, \, j \,\, {\rm odd}}t_j \Omega_j (\iota P)\Bigr )}
\\ \\
\displaystyle{
\phantom{aaaaaaaaaa}\times \,
\frac{\theta \Bigl (\vec A(\iota P)+\sum\limits_{j\geq 1, \, j \,\, {\rm odd}}\vec U_j t_j -
\vec A({\cal D}) -\vec K\Bigr )\theta \Bigl (\vec A({\cal D}) +
\vec K-\vec A(P_{\infty})\Bigr )}{\theta \Bigl (\vec A(\iota P)-
\vec A({\cal D}) -\vec K\Bigr )\theta \Bigl (
\sum\limits_{j\geq 1, \, j \,\, {\rm odd}}\vec U_j t_j
-\vec A({\cal D}) -
\vec K+\vec A(P_{\infty})\Bigr ) }}
\\ \\
\displaystyle{
=\Psi_{BA}^{\dag} (t_1, 0, t_3, 0, \ldots , P)=
\exp \Bigl (-\! \sum_{j\geq 1, \, j \,\, {\rm odd}}t_j \Omega_j (P)\Bigr )}
\\ \\
\displaystyle{
\phantom{aaaaa}\times \,
\frac{\theta \Bigl (\vec A(P)-\sum\limits_{j\geq 1, \, j \,\, {\rm odd}}\vec U_j t_j +
\vec A({\cal D}) +\vec K -2\vec A(P_{\infty})\Bigr )\theta \Bigl (\vec A({\cal D}) +
\vec K-\vec A(P_{\infty})\Bigr )}{\theta \Bigl (\vec A(P)+
\vec A({\cal D}) +\vec K-2\vec A(P_{\infty})\Bigr )
\theta \Bigl (\sum\limits_{j\geq 1, \, j \,\, {\rm odd}}\vec U_j t_j
-\vec A({\cal D}) -
\vec K+\vec A(P_{\infty})\Bigr ) }}.
\end{array}
\eeq
The tau-function of the CKP hierarchy is the square root of
\beq\label{tau1b}
\tau^{\rm KP} ({\bf t}_{\rm o})=\exp \Bigl ( -
\frac{1}{2}\sum_{i,j\geq 1,\, i,j \,\, {\rm odd}}\Omega_{ij}t_it_j\Bigr )\,
\theta \Bigl (\sum_{j\geq 1, \, j \,\, {\rm odd}}\vec U_j t_j -\vec A({\cal D})-\vec K
+\vec A(P_{\infty}) \Bigr ),
\eeq
where the divisor ${\cal D}$ satisfies the condition (\ref{iota1}).

The statement of the following theorem is in fact a corollary of 
Theorem \ref{theorem:exist} and the above identification of the 
square root of (\ref{tau1b}) with the tau-function of the CKP hierarchy but below we give its closed algebraic-geometrical proof.
\begin{theorem}\label{thm-main}
Let  $\Gamma$ be a genus $g$ smooth curve with holomorphic involution $\iota$ having at least one fixed point $P_\infty$ and  let $Y$ be the locus in the Jacobian
\beq\label{locus}
Y\subset Jac(\Gamma)=\{\vec Z\in Y|\, \vec Z+ \iota(\vec Z)=-2\vec A(P_{\infty})\}
%-{\cal K}-2\vec K\}
\eeq
Then for any point $Q\in \Gamma$ and $\vec Z\in Y$ the equation
\beq\label{th1}
\hspace{-1cm}
\begin{array}{c}
\theta \Bigl (\vec Z\Bigr )\p_1 \theta \Bigl (\vec A(Q)\! -\! \vec A(\iota Q)+\vec Z\Bigr )-
\theta \Bigl (\vec A(Q)\! -\! \vec A(\iota Q)+\vec Z\Bigr )\p_1 \theta \Bigl (\vec Z\Bigr )
\\ \\
\phantom{aaaaaaaaaaaaaaaa}+\,
2\Omega_1(Q)\theta \Bigl (\vec Z\Bigr )\theta \Bigl (\vec A(Q)\! -\!
\vec A(\iota Q)+\vec Z \Bigr)=
C(Q) \theta^2 \Bigl (\vec A(Q)+\vec Z\Bigr )
\end{array}
\eeq
with
$$\p_1\theta (\vec Z):= \p_t \theta (\vec Z +\vec U_1 t)\Bigr |_{t=0}$$
holds.
\end{theorem}
\noindent
{\bf Remark.} Note that $Y$ is the locus of vectors such that
$\vec Z=-\vec A({\cal D})-\vec K$, where the divisor ${\cal D}$ satisfies
the condition (\ref{iota1}).

\noindent
{\it Proof.} Let us fix a point $Q\in \Gamma$, an effective divisor ${\cal D}$ of degree $g$ and
define the auxiliary Baker-Akhiezer function $\Psi_{Q}({\bf t}_{\rm o}, P)$
by the following properties:
\begin{itemize}
\item[$1^0$.] Outside $P_{\infty}$ the singularities of $\Psi_Q$ are poles at the divisor
${\cal D}+\iota Q$;
\item[$2^0$.] It has simple zero at the point $Q$,
i.e., $\Psi_{Q}({\bf t}_{\rm o}, Q)=0$;
\item[$3^0$.] In a small neighborhood of $P_{\infty}$ the function $\Psi_Q$ has the form
\beq\label{th3}
\Psi_{Q}({\bf t}_{\rm o}, P)=e^{\zeta ({\bf t}_{\rm o}, k)}\Biggl (
1+\sum_{j\geq 1}\xi_{j, Q}({\bf t}_{\rm o})k^{-j}\Biggr ), \qquad k=k(P).
\eeq
\end{itemize}
The standard argument shows that this function is unique up to a common factor.
The explicit formula for $\Psi_Q$ in theta-functions is
\beq\label{th2}
\Psi_Q({\bf t}_{\rm o}, P)=\frac{\theta \Bigl (\vec A(P)
-\vec A(\iota Q)+\vec A(Q)+\vec Z_{\bf t_{\rm o}}\Bigr )\theta \Bigl (\vec Z\Bigr )}{\theta
\Bigl (\vec A(Q)-\vec A(\iota Q)+\vec Z_{\bf t_{\rm o}}\Bigr )\theta \Bigl (
\vec A(P)+\vec Z\Bigr )}
\, \exp \Biggl (\Omega_0(P)+\!\!\!\sum_{j\geq 1, \, {\rm odd}}t_j \Omega_j(P)\Biggr ),
\eeq
where $\displaystyle{\vec Z_{\bf t_{\rm o}}=\vec Z +\sum\limits_{j\geq 1, \, {\rm odd}}
U_jt_j}$ and $\Omega_0$ is the abelian integral of the normalized dipole differential
$d\Omega_0$ with simple poles at the points $Q, \iota Q$ with residues $\pm 1$:
$$
\Omega_0(P)=\int_{Q_0}^P d\Omega_0.
$$

\medskip \noindent
{\bf Remark.}
The standard Baker-Akhiezer function $\Psi_{BA}$ corresponds to the case $Q=P_{\infty}$.

\medskip

Consider the differential $\widetilde{d\Omega} (P)= \p_{t_1}\!\Psi_Q(P) \Psi_Q(\iota P)d\Omega (P)$,
where $d\Omega$ is the differential entering the bilinear relation (\ref{ba701}).
It is a meromorphic differential on $\Gamma$ with the only pole at $P_{\infty}$. Hence it has no
residue $P_{\infty}$. Computing the residue in terms of the coefficients of the expansion
(\ref{th3}), we get
\beq\label{th4}
2\xi_{2,Q}-\xi^2_{1,Q}+\p_{t_1} \xi_{1, Q}+c_1=0,
\eeq
where $c_1$ is a constant defined by the Laurent expansion of $d\Omega$ at $P_{\infty}$.

Consider now the differential
$d\Omega_Q (P)= \Psi_Q(P) \Psi_{BA} (\iota P)d\Omega (P)$. It is a meromorphic differential
with poles at $P_{\infty}$ and $\iota Q$. Therefore,
\beq\label{th5}
f_Q:=\res_{P_{\infty}}d\Omega_Q =\xi_{1, Q}-\xi_1 =-\res_{\iota Q}d\Omega_Q =-\phi_Q \phi,
\eeq
where
\beq\label{th6}
\phi_Q:= \res_{\iota Q}(\Psi_Q d\Omega ), \qquad
\phi = \Psi_{BA} ({\bf t}_{\rm o}, \iota Q).
\eeq

The residue argument for the differential $\widetilde{d\Omega}_Q(P)=
\p_{t_1}\!\Psi_Q(P) \Psi_{BA} (\iota P)d\Omega (P)$ gives the relation
\beq\label{th7}
\xi_{2, Q}+\xi_2 -\xi_{1, Q}\xi_1 +\p_{t_1}\xi_{1, Q} +c_1 =-(\p_{t_1}\phi_Q)\phi.
\eeq
Then, using (\ref{th4}), we obtain
\beq\label{th8}
\frac{1}{2}\, (f_Q^2 +\p_{t_1}f_Q)=-(\p_{t_1}\phi_Q)\phi .
\eeq
From comparison of (\ref{th5}) and (\ref{th8}) it follows that
\beq\label{th9}
\p_{t_1}\log \phi_Q =\frac{1}{2}\, (f_Q +\p_{t_1}\log f_Q).
\eeq

Recalling the definition of $\phi_Q$ and using formula (\ref{th2}), we get
\beq\label{th10}
\p_{t_1}\log \phi_Q=\p_{t_1}\log \left (
\frac{\theta \Bigl (\vec A(Q)+\vec Z_{{\bf t}_{\rm o}}\Bigr )}{\theta \Bigl (
\vec A(Q)-\vec A (\iota Q)+\vec Z_{{\bf t}_{\rm o}}\Bigr )}\right )+\Omega_1(\iota Q).
\eeq
The expansion of (\ref{th2}) around $P_{\infty}$ yields
\beq\label{th11}
f_Q=\p_{t_1}\log \left (
\frac{\theta \Bigl (\vec Z_{{\bf t}_{\rm o}}\Bigr )}{\theta \Bigl (
\vec A(Q)-\vec A (\iota Q)+\vec Z_{{\bf t}_{\rm o}}\Bigr )}\right )+\Omega_{01},
\eeq
where $\Omega_{01}$ equals the coefficient at $k^{-1}$ in the expansion of $\Omega_0$ at
$P_{\infty}$. The Riemann's  bilinear relation for the differentials
$d\Omega_1$ and $d\Omega_0$ has the form
\beq\label{th12}
\Omega_{01}=\Omega_1(\iota Q)-\Omega_1(Q)=2\Omega_1(\iota Q).
\eeq
Therefore, equations (\ref{th8}) and (\ref{th11}) imply
\beq\label{th13}
\p_{t_1}\log \left (
\frac{\theta^2 \Bigl (\vec A(Q)+\vec Z_{{\bf t}_{\rm o}}\Bigr )}{\theta \Bigl (
\vec A(Q)-\vec A (\iota Q)+\vec Z_{{\bf t}_{\rm o}}\Bigr )
\theta \Bigl (\vec Z_{{\bf t}_{\rm o}}\Bigr )}\right )=
\p_{t_1}\log f_Q.
\eeq
Equation (\ref{th11}) and (\ref{th13}) with ${\bf t}_{\rm o}=0$
after integration in $t_1$ give (\ref{th1}) with constant $C(Q,\vec Z)$
which is $\p_{t_1}$-invariant, i.e. $C(Q,\vec Z)=C(Q,\vec Z+t_1\vec U_1)$ for
any value of $t_1$. For a generic curve the complex line $\vec Z+t_1\vec U_1$ is
dense in the Jacobian. Hence, the integration constant $C$ does not depend on
$\vec Z$ and depends on $Q$ only.
Since the matrix of $b$-periods depends analytically on the
curve and $C$ is independent of $\vec Z$ for generic curve it is independent of 
$\vec Z$ for any curve.
\square

\subsection{Degeneration of algebraic-geometrical solutions: soliton solutions}

The algebraic-geometrical integration scheme naturally extends to the case of singular curves.
In particular, the case when $\Gamma$ is the Riemann sphere $\CC P^1$ with nodes (double points)
corresponds to soliton solutions. $N$-soliton solutions of the CKP hierarchy are obtained by imposing certain constraints on the parameters of $2N$-soliton solutions to the KP hierarchy. We recall that
$\tau =\sqrt{\vphantom{B^{a^a}}\tau^{\rm KP}}$, with ``even'' times $t_{2k}$ put equal to zero
and it is implied that the parameters of the KP tau-function $\tau^{\rm KP}$
are chosen in a special way.

$M$-solutions of the KP hierarchy are constructed starting from a singular curve which is
$\CC P^1$ with $M$ double points. Let $z$ be the global coordinate. The Baker-Akhiezer
function has simple poles at $M$ points $q_i$. It has the form
\beq\label{sol1}
\Psi^{\rm KP}({\bf t}, z)=\exp \Bigl (\sum_{j\geq 1}t_jz^j\Bigr )
\Biggl (1+\sum_{l=1}^M \frac{y_l({\bf t})}{z-q_l}\Biggr ).
\eeq
Let us impose $M$ linear conditions of the form
\beq\label{sol2}
\res_{z=q_i}\Bigl [
\Psi^{\rm KP}({\bf t}, z)dz \Bigr ]=
-\alpha_i (p_i-q_i)\Psi^{\rm KP}({\bf t}, p_i), \quad i=1, \ldots , M,
\eeq
which mean that the points $p_i, q_i$ are glued together forming a double point.
Here $\alpha_i$ are complex parameters. These conditions make the Baker-Akhiezer
function unique (up to a common multiplier). The conditions (\ref{sol2}) are equivalent
to the following linear system for $y_l$:
\beq\label{sol3}
y_i+\sum_{l=1}^M \frac{\tilde \alpha_i y_l}{p_i-q_l}=-\tilde \alpha_i,
\eeq
where
$$
\tilde \alpha_i= \alpha_i (p_i-q_i)\exp \Bigl (\sum_{j\geq 1}(p_i^j -q_i^j)t_j\Bigr ).
$$
Solving this system, we obtain the Baker-Akhiezer function in the explicit form:
\beq\label{sol4}
\Psi^{\rm KP}=\frac{\phantom{a}\left | \begin{array}{ccccc}
1& \frac{1}{z-q_1} & \frac{1}{z-q_2} & \ldots & \frac{1}{z-q_M}
\\ \\
\tilde \alpha_1 & 1\! +\!\frac{\tilde \alpha_1}{p_1-q_1} &\frac{\tilde \alpha_1}{p_1-q_2}&
\ldots & \frac{\tilde \alpha_1}{p_1-q_M}
\\ \\
\tilde \alpha_2 & \frac{\tilde \alpha_2}{p_2-q_1} &1\! +\! \frac{\tilde \alpha_2}{p_2-q_2}&
\ldots & \frac{\tilde \alpha_2}{p_2-q_M}
\\ \ldots & \ldots & \ldots & \ldots & \ldots
\\ \\
\tilde \alpha_M & \frac{\tilde \alpha_M}{p_M-q_1} &\frac{\tilde \alpha_M}{p_M-q_2}&
\ldots & 1\! +\! \frac{\tilde \alpha_M}{p_M-q_M}
\end{array}
\right |\phantom{a}}{\left | \begin{array}{cccc}
\vphantom{\frac{A^{a^a}}{a}}1\! +\! \frac{\tilde \alpha_1}{p_1-q_1} &\frac{\tilde \alpha_1}{p_1-q_2}&
\ldots & \frac{\tilde \alpha_1}{p_1-q_M}
\\ \\
\frac{\tilde \alpha_2}{p_2-q_1} &1\! +\! \frac{\tilde \alpha_2}{p_2-q_2}&
\ldots & \frac{\tilde \alpha_2}{p_2-q_M}
\\ \ldots & \ldots & \ldots & \ldots
\\ \\
\frac{\tilde \alpha_M}{p_M-q_1} &\frac{\tilde \alpha_M}{p_M-q_2}&
\ldots & 1\! +\! \frac{\tilde \alpha_M}{p_M-q_M}
\end{array}
\right |}\exp \Bigl (\sum_{j\geq 1}t_jz^j\Bigr ).
\eeq
The denominator of this expression is the tau-function.

The general KP tau-function for $M$-soliton solution has $3M$ arbitrary parameters
$\alpha_i$, $p_i$, $q_i$ ($i=1, \ldots , M$) and is given by
\beq\label{ms1}
%\begin{array}{c}
%\displaystyle{
\tau^{\rm KP} (x, {\bf t})
%\\ \\
%\displaystyle{
=\det_{1\leq i,j\leq M}\left ( \delta_{ij}+
\alpha_i \, \frac{p_i-q_i}{p_i-q_j}\, \exp
\Bigl ((p_i-q_i)x+\!\! \sum_{k\geq 1}(p_i^k-q_i^k)t_k \Bigr )
\right ).
%\end{array}
\eeq
Let us denote this tau-function as
$$
\tau^{\rm KP} \left [ \begin{array}{c}\alpha_1 \\ p_1, q_1\end{array};
\begin{array}{c}\alpha_2 \\ p_2, q_2\end{array};
\begin{array}{c}\alpha_3 \\ p_3, q_3\end{array};
\begin{array}{c}\alpha_4 \\ p_4, q_4\end{array}; \, \cdots \, ;
\begin{array}{c}\alpha_{2N-1} \\ p_{M-1}, q_{M-1}\end{array};
\begin{array}{c}\alpha_{2N} \\ p_{M}, q_{M}\end{array}\right ].
$$
The parameters $p_i, q_i$ are sometimes called momenta of solitons.

In the CKP case we have the involution $z\to -z$ which means that the double points
should be symmetric under the involution.
The multi-soliton tau-function of the CKP hierarchy is the square root of the $\tau^{\rm KP}$
specialized as
\beq\label{ms2}
\begin{array}{c}
\displaystyle{
\tau^{\rm KP} \left [ \begin{array}{c}\alpha_0 \\ p_0, -p_0\end{array};
\begin{array}{c}\alpha_1 \\ p_1, -q_1\end{array};
\begin{array}{c}\alpha_1 \\ q_1, -p_1\end{array};
\begin{array}{c}\alpha_2 \\ p_2, -q_2\end{array};
\begin{array}{c}\alpha_2 \\ q_2, -p_2\end{array}; \, \cdots \, ;
\begin{array}{c}\alpha_{N} \\ p_{N}, -q_{N}\end{array};
\begin{array}{c}\alpha_{N} \\ q_{N}, -p_{N}\end{array}\right ]},
\end{array}
\eeq
where it is assumed that even times evolution is suppressed ($t_{2k}=0$ for all $k\geq 1$).
Clearly, the total number of independent parameters is $3N+2$. If $\alpha_0=0$, the tau-function
(\ref{ms2}) reduces to
\beq\label{ms3}
\begin{array}{c}
\displaystyle{
\tau^{\rm KP} \left [
\begin{array}{c}\alpha_1 \\ p_1, -q_1\end{array};
\begin{array}{c}\alpha_1 \\ q_1, -p_1\end{array};
\begin{array}{c}\alpha_2 \\ p_2, -q_2\end{array};
\begin{array}{c}\alpha_2 \\ q_2, -p_2\end{array}; \, \cdots \, ;
\begin{array}{c}\alpha_{N} \\ p_{N}, -q_{N}\end{array};
\begin{array}{c}\alpha_{N} \\ q_{N}, -p_{N}\end{array}\right ]},
\end{array}
\eeq
and it is this tau-function which is usually
called the $N$-soliton CKP tau-function in the literature (see, e.g. \cite{DJKM81}).
It is a specialization of $2N$-soliton KP tau-function and
has $3N$ free parameters.

The simplest example is one-soliton solution.
The tau-function for one CKP soliton is the square root
of a specialization of 2-soliton tau-function
of the KP hierarchy:
\beq\label{e2}
\tau^{\rm KP}= 1+2\alpha w-
\frac{\alpha^2 (p-q)^2}{4pq} \, w^2,
\eeq
where
\beq\label{e5}
w=e^{(p+q)x+ \zeta ({\bf t}_{\rm o}, p)+\zeta ({\bf t}_{\rm o}, q)}, \quad
\mbox{$\zeta ({\bf t}_{\rm o}, z)$ is given by (\ref{ckp7}).}
\eeq
A direct calculation shows that $\p_x \psi^2$ (where $\psi$
is given by (\ref{e3})) for the solution (\ref{e2})
is a full square for all $z$.

\medskip
\noindent
{\bf Remark.} It is instructive to prove directly that  the tau-functions (\ref{ms2}) and (\ref{ms3}) satisfy equation (\ref{ch6}).
Consider (\ref{ms3}) first. We represent the tau-function as
$$
\tau^{\rm KP}=\det_{2N\times 2N} (I+HK),
$$
where $H$ is the diagonal matrix $W_{jk}=\delta_{jk}W_j$ with matrix elements
$$
H_{2i-1}=\alpha_i (p_i+q_i)\exp \Biggl ( (p_i+q_i)x +\sum_{k\geq 1}t_k
(p_i^k -(-q_i)^k)\Biggr ), \quad i=1, \ldots , N,
$$
$$
H_{2i}=\alpha_i (p_i+q_i)\exp \Biggl ( (p_i+q_i)x +\sum_{k\geq 1}t_k
(q_i^k -(-p_i)^k)\Biggr ), \quad i=1, \ldots , N,
$$
and $K$ is the Cauchy matrix $K_{jk}=1/(x_j-y_k)$ with
$x_{2i-1}=-y_{2i}=p_i$, $x_{2i}=-y_{2i-1}=q_i$, $i=1, \ldots , N$.
We have:
$$
\p_{t_{2m}}\log \tau^{\rm KP}\Biggr |_{t_{2k}=0}=
\p_{t_{2m}}\log \det (I+HK)\Biggr |_{t_{2k}=0}=
\p_{t_{2m}}\mbox{tr}\, \log \, (I+HK)\Biggr |_{t_{2k}=0}
$$
$$
=\mbox{tr} \Bigl [ VHK(I+HK)^{-1}\Bigr ] =\mbox{tr}\, V -
\mbox{tr} \Bigl [ V(I+HK)^{-1}\Bigr ],
$$
where $V$ is the diagonal matrix $V_{jk}=\delta_{jk}V_j$ with the matrix elements
$$V_{2i-1}=-V_{2i}=p_i^{2m}-q_i^{2m}. $$
Note also that when all even times are put equal to zero, we have also
$H_{2i-1}=-H_{2i}$.
Obviously, $\mbox{tr} V=0$. A careful inspection shows that
$(I+HK)^{-1}_{2i-1, 2i-1}=(I+HK)^{-1}_{2i, 2i}$, and, therefore,
$\mbox{tr} \Bigl [ V(I+HK)^{-1}\Bigr ]=0$, too, and the conditions
(\ref{ch6}) are satisfied. Indeed, permuting rows and columns, one can see that
the diagonal $(2i-1, 2i-1)$ and $(2i, 2i)$ minors of the matrix
$I+HK$ are equal.
As for the tau-function (\ref{ms2}) with $\alpha_0\neq 0$, it is obvious that
the additional pair of soliton momenta of the form $p_0, -p_0$ does not lead to
any extra dependence on the even times, and so the conditions (\ref{ch6}) are still
satisfied.

\section{Elliptic solutions}

By elliptic solutions of the CKP equation (\ref{ckp0}) we mean
solutions $u$ that are double-periodic in the complex plane of the variable $x$
with periods $2\omega_1, 2\omega_2, \, {\rm Im}(\omega_2/ \omega_1)>0$.
Equations of motion for their poles and their algebraic integrability is an
easy corollary of the established above relation between the CKP and KP hierarchies
and the well-developed theory of elliptic solutions to the KP hierarchy,
equivalent to the theory of the elliptic Calogero-Moser (eCM) system.

Namely, elliptic solutions of the CKP equation can be
extended to elliptic solutions of the KP equation and further to the whole KP hierarchy. From that perspective the pole dynamics of the elliptic solutions of the CKP equation in $t_3$ is just the restriction of $t_3$-dynamics generated
by the Hamiltonian $H_3$ of the eCM system,
\beq\label{ca8}
\left \{
\begin{array}{l}
\displaystyle{\dot x_i =-3p_i^2 +3 \sum_{j\neq i}\wp (x_i-x_j)-6c}
\\ \\
\displaystyle{\dot p_i =-3\sum_{j\neq i}(p_i+p_j)\wp '(x_i-x_j)},
\end{array}
\right.
\eeq
onto the locus of turning points $p_i=0$ that is invariant under
${\bf t_{\rm o}}$ flows of the eCM system, i.e.
\beq\label{ca1a}
\dot x_i =3\sum_{k\neq i}\wp (x_i-x_k)-6c,
\eeq
where $c$ is a constant and dot means the $t_3$-derivative.
Here $\wp$ is the Weiershtrass $\wp$-function which is an
even double-periodic function with periods $2\omega_1, \, 2\omega_2$
having second order poles at the lattice points
$2\omega_1 m_1+2\omega_2 m_2$ with integer $m_1, m_2$ and
$$
\wp (x)=\frac{1}{x^2} + O(x^2), \quad x\to 0.
$$

For further use recall the definitions of the Weierstrass functions. The Weierstrass
$\sigma$-function is given by the infinite product
$$
\sigma (x)=\sigma (x |\, \omega_1 , \omega_2)=
x\prod_{s\neq 0}\Bigl (1-\frac{x}{s}\Bigr )\, e^{\frac{x}{s}+\frac{x^2}{2s^2}},
\ \  s=2\omega_1 m_1+2\omega_2 m_2\,, \ \ m_1, m_2\in \ZZ.
$$
The Weierstrass $\zeta$- and $\wp$-functions are connected with the $\sigma$-function
as follows: $\zeta (x)=\sigma '(x)/\sigma (x)$,
$\wp (x)=-\zeta '(x)=-\p_x^2\log \sigma (x)$.

\medskip
The algebraic integrability of the eCM system established
in \cite{Krichever80} restricted to the locus of turning points can be stated as follows.

\begin{theorem} For each set of constants $x_i^0\neq x_j^0$
define the algebraic curve $\Gamma$ by the characteristic equation  $\det (zI-L)=0$ for the matrix
\beq\label{laxckp} L_{ii}=0,\qquad L_{ij}=-\Phi(x_i^0-x_j^0,\lambda), \qquad i\neq j,
\eeq
where
\beq\label{phidef}
\Phi (x, \lambda )=\frac{\sigma (x+\lambda )}{\sigma (\lambda )\sigma (x)}\,
e^{-\zeta (\lambda )x}.
\eeq
Let $P_\infty$ be the point on $\Gamma$ that is the
pre-image of $\lambda=0$ in  the neighborhood of which $z$ has the
expansion $z=-(n-1) \lambda^{-1}+O(\lambda )$.
Then the solution of (\ref{ca1a})
with the initial conditions $x_i(0)=x_i^0$ are roots $x_i(t_3)$ of the equation
\beq\label{algformula}
\theta \Bigl(\vec U_1 x_i+\vec U_3 t_3+ \vec Z\,\Bigl |\,T\Bigr )=0 .
\eeq
Here $\theta(z\,|\,T)$ is  Riemann theta-function defined by the matrix of $b$-periods of normalized holomorphic differentials on $\Gamma$; the vectors $\vec U_j$ are given
by (\ref{qp5}) with $d\Omega_j$ defined in (\ref{defdomega}); the vector $\vec Z$ is in the
locus $Y$ defined in (\ref{locus}), where the involution $\iota$ of the Jacobian is induced by the involution $\iota(z,\lambda)\to (-z,-\lambda)$ of $\Gamma$.
\end{theorem}

The elliptic solutions are particular cases of the general algebraic-geometrical
solutions considered in the
previous section. The corresponding spectral data are singled
out by the following constraint: the vectors $2\omega_1\vec U_1, 2\omega_2 \vec U_1$
are in the lattice of
periods of the Jacobian of the spectral curve, where
$\vec U_1$ is the vector of $b$-periods of the normalized
differential with the only pole (of order 2) at the marked point $P_\infty$.

\subsection{The generating problem}

For completeness, in this section we present the scheme proposed in
\cite{Krichever80} which allows one to derive the
equations of motion for poles of elliptic solutions to a variety of
soliton equations together with their Lax-type representation
(see more in \cite{kr-nested}). With the help of this scheme
we will get the equations (\ref{ca1a}) and the Lax matrix (\ref{laxckp})
directly without use of relations to the theory of the eCM system.

The elliptic solution of the CKP equation is an
elliptic function with double poles at the points $x_i$:
\beq\label{int7}
u=-\frac{1}{2}\sum_{i=1}^{n}\wp (x-x_i)+c,
\eeq
where $c$ is a constant. The poles depend on the times $t_3$, $t_5$ (as well as on the higher times)
and are assumed to be
all distinct. The corresponding CKP tau-function has the form
\beq\label{int7a}
\tau (x, {\bf t}_{\rm o})=C_0e^{cx^2/2}\left ( \prod_{i=1}^n
\sigma (x-x_i({\bf t}_{\rm o}))\right )^{1/2}.
\eeq

In the rest of this section we denote $t_3=t$.
According to the scheme proposed in \cite{Krichever80},
the basic tool is the auxiliary linear problem
$\p_{t}\Psi =B_3\Psi$ for the wave
function $\Psi$, i.e.,
\beq\label{ba0}
\p_t \Psi =\p_x^3\Psi +6u \p_x \Psi +3u'\Psi ,
\eeq
for which one can state the following problem: characterize
an elliptic in $x$ function $u$ of the form (\ref{int7}) for which equation ({\ref{ba0})
has {\it double-Bloch solutions} $\Psi (x)$, i.e., solutions such that
$\Psi (x+2\omega_{\alpha} )=B_{\alpha} \Psi (x)$
with some Bloch multipliers $B_{\alpha}$}.
Equations (\ref{e3}), (\ref{e4}) imply that the
wave function has simple poles at the points $x_i$.
Therefore, if a double-Bloch solution exists, then it is of the following pole ansatz form:
\beq\label{ba1}
\Psi = e^{xz+tz^3}\sum_{i=1}^n c_i \Phi (x-x_i, \lambda ),
\eeq
where the coefficients $c_i$ do not depend on $x$ (but do depend on $t$, $z$ and $\lambda$).
Indeed, the function $\Phi (x, \lambda )$ given by formula (\ref{phidef})
has the following monodromy properties:
\beq\label{phimon}
\Phi (x+2\omega_{\alpha} , \lambda )=e^{2(\zeta (\omega_{\alpha} )\lambda -
\zeta (\lambda )\omega_{\alpha} )}
\Phi (x, \lambda ), \quad \alpha =1,2.
\eeq
Therefore, the wave function $\Psi$ given by (\ref{ba1})
is a double-Bloch function with Bloch multipliers
$B_{\alpha}=e^{2(\omega_{\alpha} z + \zeta (\omega_{\alpha} )\lambda -
\zeta (\lambda )\omega_{\alpha} )}$ parameterized by $z$ and $\lambda$.

In what follows we will often suppress the second argument of $\Phi$ writing simply
$\Phi (x)=\Phi (x, \lambda )$. For further use note also that $\Phi$ has a simple pole at $x=0$ with residue $1$. The coefficients $\beta_{1}, \beta_2$ of its expansion
$$
\Phi (x, \lambda )=\frac{1}{x}+\beta_1 x +\beta_2 x^2 +O(x^3) \quad
\mbox{as $x\to 0$},
$$
are equal to
\beq\label{alpha}
\beta_1 =-\frac{1}{2} \, \wp (\lambda ), \quad
\beta_2 =-\frac{1}{6} \, \wp '(\lambda ).
\eeq
The function $\Phi$
We will also need the $x$-derivatives
$\Phi '(x, \lambda )=\p_x \Phi (x, \lambda )$, $\Phi ''(x, \lambda )=\p^2_x \Phi (x, \lambda )$
and so on.

\begin{theorem}
The equations of motion (\ref{ca1a}) for
poles $x_i$ of elliptic solutions as functions of $t=t_3$ have the following
commutation representation of the Manakov's triple kind:
%\beq\label{ca1a}
%\dot x_i =3\sum_{k\neq i}\wp (x_i-x_k)-6c,
%\eeq
%where $c$ is a constant and
%dot means the $t$-derivative.
\beq\label{ca6}
\dot L+[L,M]=3D'(zI-L),
\eeq
where
\beq\label{laxckp1} L_{ii}=0,\qquad L_{ij}=-\Phi(x_i-x_j,\lambda), \qquad i\neq j;
\eeq
the matrix $M$ is defined by (\ref{ca3}), and
$D'$ is the diagonal matrix $\displaystyle{D'_{ik}=
\delta_{ik}\sum_{j\neq i}\wp '(x_i-x_j)}$.
\end{theorem}

\noindent
{\it Proof.}
Substituting (\ref{ba1}) into (\ref{ba0}) with
$u=-\frac{1}{2}\displaystyle{\sum_{i}\wp (x-x_i)}+c$,
we get:
$$
\sum_i \dot c_i\Phi (x-x_i)-\sum_i c_i \dot x_i \Phi '(x-x_i)=3z^2
\sum_i c_i \Phi '(x-x_i)+3z\sum_i c_i \Phi ''(x-x_i)+\sum_i c_i \Phi '''(x-x_i)
$$
$$
-3z\Bigl (\sum_k \wp (x-x_k)\Bigr ) \Bigl (\sum_i c_i \Phi (x-x_i)\Bigr )
-3\Bigl (\sum_k \wp (x-x_k)\Bigr ) \Bigl (\sum_i c_i \Phi ' (x-x_i)\Bigr )
$$
$$
-\frac{3}{2}\Bigl (\sum_k \wp '(x-x_k)\Bigr ) \Bigl (\sum_i c_i \Phi (x-x_i)\Bigr )
+6cz\sum_i c_i\Phi (x-x_i) +6c\sum_i c_i \Phi ' (x-x_i).
$$
It is enough to cancel all poles in the fundamental domain
which are at the points $x_i$ (up to fourth order).
It is easy to see that poles of the fourth order cancel identically.
A direct calculation shows that the conditions of cancellation of third, second and first
order poles have the form
\beq\label{ba2a}
zc_i=-\sum_{k\neq i}c_k \Phi (x_i-x_k),
\eeq
\beq\label{ba2}
c_i\dot x_i=-3z^2c_i +3c_i \sum_{k\neq i}\wp (x_i-x_k)-3z \sum_{k\neq i}c_k \Phi (x_i-x_k)
-6cc_i,
\eeq
\beq\label{ba3}
\begin{array}{lll}
\dot c_i &=&\displaystyle{
-3(\beta_1 z+\beta_2)c_i -3zc_i\sum_{k\neq i}\wp (x_i-x_k)
+\frac{3}{2}\, c_i \sum_{k\neq i}\wp '(x_i-x_k)}
\\ &&\\
&&\phantom{aaaaaaa}\displaystyle{-\, 3z \sum_{k\neq i}c_k \Phi '(x_i-x_k)
-\frac{3}{2}\sum_{k\neq i}c_k\Phi ''(x_i-x_k)+6czc_i}
\end{array}
\eeq
which have to be valid for all $i=1, \ldots , n$.
Substitution of (\ref{ba2a}) into (\ref{ba2}) gives (\ref{ca1a})
(if the coefficients $c_i$ are not identically zero).
The conditions (\ref{ba2a}), (\ref{ba3}) can be rewritten in the matrix form as
linear problems for a vector ${\bf c} =(c_1, \ldots , c_n)^T$:
\beq\label{ca2}
\left \{ \begin{array}{l}
L{\bf c} = z{\bf c}
\\ \\
\dot {\bf c} =M{\bf c},
\end{array}
\right.
\eeq
where $L$ is the matrix (\ref{laxckp1}),
\beq\label{ca3}
\begin{array}{l}
M= -3(\beta_1 z+\beta_2-2cz)I -3zB -3zD -\frac{3}{2}\, C +\frac{3}{2}\, D'
\end{array}
\eeq
and the $n\! \times \! n$ matrices $I$, $B$, $C$, $D$, are given by
$I_{ik}=\delta_{ik}$,
\beq\label{mat}
\begin{array}{l}
B_{ik}=(1-\delta_{ik})\Phi ' (x_i-x_k),
\\ \\
C_{ik}=(1-\delta_{ik})\Phi '' (x_i-x_k),
\\ \\
\displaystyle{D_{ik}=\delta_{ik}\sum_{j\neq i}\wp (x_i-x_j),}
\end{array}
\eeq
The matrices $L, B,C$ are off-diagonal while the matrices $D, D'$ are diagonal.

The linear system (\ref{ca2}) is overdetermined.
Differentiating the first equation in (\ref{ca2}) with respect to $t$, we see that
the compatibility condition of the linear problems (\ref{ca2}) is
\beq\label{ca4}
\Bigl (\dot L+[L,M]\Bigr ) {\bf c} =0.
\eeq
One can prove the following matrix identity (see the appendix):
\beq\label{ca5}
\dot L+[L,M]=3D'(zI-L) -[\dot X \! -\! 3D, \, B],
\eeq
where $X$ is the diagonal matrix $X_{ik}=\delta_{ik}x_i$.
Since $(zI-L){\bf c}=0$ according to (\ref{ba2a}) and $\dot X =3D-6cI$ according to
(\ref{ca2}), we see from (\ref{ca5})
that the compatibility condition (\ref{ca4}) is satisfied. From (\ref{ca5}) it follows that
the equations of motion have the commutation representation
of the Manakov's triple kind (\ref{ca6}) \cite{Manakov}.
\square

\subsection{The integrals of motion and the spectral curve}

It follows from equation (\ref{ca6}) that the characteristic polynomial of the matrix $L$
is an integral of motion. Indeed,
\beq\label{ca7}
\begin{array}{c}
\displaystyle{\frac{d}{dt}\, \log \det  (L-zI )=
\frac{d}{dt}\, \mbox{tr}\log (L-zI )}
\\ \\
\displaystyle{=\, \mbox{tr}\Bigl [ \dot L (L-zI )^{-1}\Bigr ]=
-3\, \mbox{tr}D' =0,}
\end{array}
\eeq
where we have used equation (\ref{ca6}) and the fact that
$\displaystyle{\mbox{tr}\, D'=\sum_{i\neq j}\wp '(x_i-x_j)=0}$ ($\wp '$ is an odd function).
The expression
$
R(z, \lambda )=\det (zI-L(\lambda ) )
$
is a polynomial in $z$
of degree $n$. Its coefficients are integrals of motion
(some of them may be trivial). For example:
$$
\begin{array}{l}
n=2: \qquad \det\limits_{2\times 2}(zI-L)=z^2+\wp (x_{12})-\wp (\lambda ),
\\ \\
n=3: \qquad \det\limits_{3\times 3}(zI-L)=z^3 +z\Bigl (\wp (x_{12})+\wp (x_{13})+\wp (x_{23})-
3\wp (\lambda )\Bigr )-\wp '(\lambda ),
\end{array}
$$
where $x_{ik}\equiv x_i-x_k$.

\medskip
\noindent
{\bf Remark.}
Although the Lax equation for matrices $L,M$ does not hold, it follows from (\ref{ca7}) that
traces of the Lax matrix $L$ (and therefore its eigenvalues) are integrals of motion:
$\p_t \mbox{tr}\, L^m =0$, $m\geq 1$. (This is equivalent to the equalities
$\mbox{tr} \, (D'L^m)=0$ for $m\geq 1$ which are based on certain non-trivial identities
for the $\wp$-function.) This mans that the time evolution is an isospectral transformation
of the Lax matrix $L$. Therefore, there should exist a matrix $M_0$ such that the Lax equation
$
\dot L+[L,M_0]=0
$
holds. In order to find it explicitly, we first note that by virtue of the matrix identity
(\ref{A4}) (see the appendix) we can write equation (\ref{ca6}) in the form
$
\dot L+[L,\hat M]=-3D'L,
$
where
$$
\hat M=M+3z\Bigl ((\beta_1-2c)I+B+D\Bigr )=-3\beta_2 I -\frac{3}{2}\, (C-D')
$$
does not depend on $z$. Using again the identity (\ref{A4}), one can see that
\beq\label{lax2}
\begin{array}{c}
M_0=\hat M -3(B+D)L=-3\beta_2 I -\frac{3}{2}\, (C-D')-3(B+D)L
\\ \\
= M+3z(\beta_1 -2c)I +3(B+D) (zI-L)
\end{array}
\eeq
($\beta_1$, $\beta_2$ are given in (\ref{alpha})).

\medskip

The embedding into the Calogero-Moser dynamics discussed above
implies that the integrals of motion $I_k$ for the dynamical system (\ref{ca1a})
are restrictions of the Calogero-Moser integrals of motion to the subspace
of the phase space with $p_i=0$.
For example:
\beq\label{i4}
\begin{array}{l}
I_2=\displaystyle{\sum\limits_{i<j}\wp (x_{ij})},
\\ \\
I_4=\displaystyle{\sum\limits_{i<j<k<l}\Bigl [\wp (x_{ij})\wp (x_{kl})+\wp (x_{ik})\wp (x_{jl})+
\wp (x_{il})\wp (x_{jk})\Bigr ]}.
\end{array}
\eeq
The spectral curve $\Gamma$ is defined by the equation $R(z, \lambda )=\det (zI-L(\lambda ))=0$.
It is an $n$-sheet covering of the elliptic curve ${\cal E}$ uniformized by the variable $\lambda$
and realized as a factor
of the complex plane with respect to the lattice generated by $2\omega_1$, $2\omega_2$.
Since $L(-\lambda )=-L^T (\lambda )$, it is easy to see that
the curve $\Gamma$ is equipped with the holomorphic involution
$\iota :(z, \lambda )\to (-z, -\lambda )$.
As it was already mentioned, the equation of the spectral curve
(the characteristic equation of the Lax matrix) is an integral of motion.

\begin{proposition}(\cite{Krichever80}) For generic values of $x_i$ the spectral curve is smooth of
genus $g=n$.
\end{proposition}

\subsection{The wave function as the Baker-Akhiezer function on the spectral curve}

Let $P$ be a point of the spectral curve $\Gamma$, i.e. $P=(z, \lambda )$, where
$z$ and $\lambda$ are connected by the equation $R(z, \lambda )=0$.
To each point $P$ of the curve there corresponds a single eigenvector
${\bf c}(0,P)=(c_1(0,P), \ldots , c_n(0,P))^T$
of the matrix
$L(t=0, \lambda )$ normalized by the condition $c_1(0, P)=1$.
The non-normalized components $c_i$ are equal to
$\Delta_i (0, P)$, where $\Delta_i (0, P)$ are suitable minors of the matrix
$zI-L(0, \lambda )$. They are holomorphic functions on $\Gamma$ outside the
points above $\lambda =0$. After normalizing the first component, all other components
$c_i(0,P)$ become meromorphic functions on $\Gamma$ outside the points $P_j$ located
above $\lambda =0$. Let ${\cal D}'$ be the poles divisor of the vector $\bf c$ with coordinates $c_i$.
Unlike the spectra curve which is time-independent the divisor ${\cal D}'$ depends on the initial data.

\begin{lemma} \label{lem25}The sum of the divisors ${\cal D}'$
and $\iota({\cal D}')$ is the zero divisor of a holomorphic
differential on the spectral curve, i.e. the equation
\beq\label{diotad}
{\cal D}'+\iota({\cal D}')={\cal K}
\eeq
holds.
\end{lemma}
\noindent
{\it Proof}. The idea of the proof goes back to the proof of Theorem 4 in \cite{Kr87}.
Taking the differential of the eigenvalue equation $(zI-L(\lambda )){\bf c}(P)=0$
and using the equation ${\bf c}^T(\iota P)(zI-L(\lambda ))=0$,
which follows from the definition of the involution, we get the equation
$$
{\bf c}^T(\iota P)(dzI-dL(\lambda )){\bf c}(P)=0,
$$
or
\beq\label{s10}
\left < {\bf c}(\iota P), {\bf c}(P)\right >dz =
\left < {\bf c}(\iota P), L_{\lambda}{\bf c}(P)\right >d\lambda,
\eeq
where $L_{\lambda}=\p L/\p \lambda$ and $\displaystyle{
\left < {\bf c}(\iota P), {\bf c}(P)\right >=\sum_i c_i(\iota P)c_i(P)}$.
For a generic initial data the spectral curve is smooth, i.e. the differentials $dz$ and $d\lambda$
have no common zeros. Then from (\ref{s10}) it follows that the zeros of
the differential $d\lambda$ (which are ramification points of the covering
$\Gamma \to {\cal E}$) coincide with the zeros of the
function $\left < {\bf c}(\iota P), {\bf c}(P)\right >$. Therefore,
the differential
\beq\label{s11}
d\Lambda = \frac{d\lambda}{\left < {\bf c}(\iota P), {\bf c}(P)\right >}
\eeq
is a holomorphic differential on the curve $\Gamma$. Its $2g-2$ zeros at the points,
where the vectors ${\bf c}(P)$ and ${\bf c}(\iota P)$ have poles.
\square

For completeness let us outline
the arguments that ultimately lead to the proof of the
algebraic integrability of equations (\ref{ca1a}).

A particular case of Theorem 2 in \cite{Krichever80} is the following statement.

\begin{theorem} The function
\beq\label{s9a}
\hat \Psi (x, t, P)=e^{-\zeta (\lambda )x_1(0)}
\sum_{i=1}^{n}c_i(t, P)\Phi (x-x_i , \lambda )e^{zx+z^3t}
\eeq
is the one-point Baker-Akhiezer function on the spectral
curve $\Gamma$ with the marked point $P_{\infty}$ (one of pre-images
of $\lambda=0$) corresponding to the divisor ${\cal D}={\cal D}'+P_\infty$.
\end{theorem}

\noindent
By definition the function $\hat \Psi$ has poles at $x_i(t)$. From the theta-functional formula (\ref{ba501}) for the Baker-Akhiezer function
it follows that $x_i$ are zeros of the second factor
in the denominator, i.e. they are roots in $x$ of the equation
$$\theta \Bigl (\vec U_1x+\vec U_3 t-\vec A({\cal D}) -
\vec K+\vec A(P_{\infty})\Bigr )=0.
$$
From Lemma \ref{lem25} it follows that the pole divisor
$\cal D$ of the Baker-Akhiezer function satisfies the equation
\beq\label{s12}
{\cal D}+\iota {\cal D}-2P_\infty ={\cal K},
\eeq
where ${\cal K}$ is the canonical class. This is precisely the condition (\ref{iota1}) on the
divisor of poles of the Baker-Akhiezer function for algebraic-geometric solutions
to the CKP equation. This completes the proof of (\ref{algformula}) since (\ref{s12}) is equivalent to equation (\ref{locus}) for the vector $\vec Z$ in (\ref{algformula}).

\subsection{Degenerations of elliptic solutions}

\subsubsection{Trigonometric solutions}

In the degenerate case, when one of the periods tends to infinity,
the elliptic solutions become trigonometric (hyperbolic). We consider trigonometric
solutions which vanish at infinity:
$$
u(x, {\bf t})=- \, \frac{1}{2}\sum_{i=1}^n
\frac{\gamma^2}{\sinh^2 (\gamma (x\! -\! x_i({\bf t}))},
$$
where $\gamma$ is a complex parameter. When $\gamma$ is purely imaginary (respectively, real),
one deals with trigonometric (respectively, hyperbolic) solutions.
The equations of motion for the poles are
\beq\label{trig1}
\dot x_i=3\sum_{k\neq i}\frac{\gamma^2}{\sinh^2(\gamma (x_i\! -\! x_k))}-\gamma^2.
\eeq
Tending the spectral parameter $\lambda$ to infinity, we find the
Lax matrix in the form
\beq\label{trig2}
L_{ij}=-\, \frac{\gamma(1-\delta_{ij})}{\sinh (\gamma (x_i-x_j))}.
\eeq
Note that it is antisymmetric.

As is shown in \cite{Z20}, the KP tau-function for trigonometric solutions
has the following determinant representation:
\beq\label{trig3}
\tau^{\rm KP}(x, {\bf t})=
\det_{n\times n}\Biggl (e^{2\gamma x}I-\exp \Bigl (-\sum_{k\geq 1}t_k {\cal L}_k\Bigr )
e^{2\gamma X_0}\Biggr )=\prod_{j=1}^n (e^{2\gamma x}-e^{2\gamma x_j({\bf t})}),
\eeq
where $X_0=\mbox{diag}\, (x_1(0), \ldots , x_n(0))$ and
\beq\label{trig4}
{\cal L}_k=(L_0+\gamma I)^k-(L_0-\gamma I)^k, \qquad L_0 =L({\bf t}=0).
\eeq
We see that ${\cal L}_k$ is a polynomial in $L_0$ of degree $k-1$. If $k$ is even
(respectively, odd), ${\cal L}_k$ contains only odd (respectively, even) powers of $L_0$.

It is easy to see that this tau-function satisfies the conditions (\ref{ch6}), and,
therefore, gives rise to the CKP tau-function
\beq\label{trig5}
\tau (x, {\bf t}_{\rm o})=\Biggl (\det_{n\times n}
\Bigl (e^{2\gamma x}I-\exp \Bigl (-\!\!\!\!\!\sum_{k\geq 1, \,\, k \,\, {\rm odd}}
\!\! t_k {\cal L}_k\Bigr )
e^{2\gamma X_0}\Bigr )\Biggr )^{1/2}.
\eeq
Indeed, we have
$$
\p_{t_{2m}}\log \tau^{\rm KP}\Biggl |_{t_{2k}=0}=
\p_{t_{2m}}\mbox{tr}\log \Biggl (e^{2\gamma x}I-\exp \Bigl (-\sum_{k\geq 1}t_k {\cal L}_k\Bigr )
e^{2\gamma X_0}\Biggr )\Biggl |_{t_{\rm e}=0}
$$
$$
=\mbox{tr} \Biggl [ {\cal L}_{2m} \exp \Bigl (-\!\!\!\!\!
\sum_{k\geq 1, \,\, k \,\, {\rm odd}}t_k {\cal L}_k\Bigr )
\Biggl (e^{2\gamma x}I-\exp \Bigl (-\!\!\!\!\!\sum_{k\geq 1, \,\, k \,\, {\rm odd}}
\!\! t_k {\cal L}_k\Bigr )
e^{2\gamma X_0}\Biggr )^{-1}\Biggr ].
$$
But this is zero for all $m\geq 1$
because $\mbox{tr}\, L_0^{2l-1}=0$ for all $l\geq 1$ and,
as it was said above, ${\cal L}_{2m}$ contains only odd
powers of $L_0$ while all other ${\cal L}_k$ in this expression contain only even
powers of $L_0$.

\subsubsection{Rational solutions}

In the most degenerate case, when both periods tend to infinity, $\wp (x)\to 1/x^2$
and the elliptic solutions become rational:
$$
u(x, {\bf t})=- \, \frac{1}{2}\sum_{i=1}^n \frac{1}{(x-x_i({\bf t}))^2}.
$$
This corresponds to the limit $\gamma \to 0$ in the trigonometric
solutions. The equations of motion for the poles are
\beq\label{rat1}
\dot x_i=3\sum_{k\neq i}\frac{1}{(x_i-x_k)^2}.
\eeq
Tending the spectral parameter $\lambda$ to infinity, $\lambda =\infty$, we find the
(antisymmetric) Lax matrix in the form
\beq\label{rat2}
L_{ij}=-\, \frac{1-\delta_{ij}}{x_i-x_j}.
\eeq

It is known that the KP tau-function for rational solutions has the following
determinant representation (see, e.g. \cite{Shiota}):
\beq\label{rat3}
\tau^{\rm KP}(x, {\bf t})=
\det_{n\times n}\Bigl (xI-X_0 +\sum_{k\geq 1}kt_k L_0^{k-1}\Bigr )=
\prod_{j=1}^n (x-x_j({\bf t})),
\eeq
where $X_0=\mbox{diag}\, (x_1(0), \ldots , x_n(0))$ and $L_0 =L({\bf t}=0)$.
It is easy to see that this tau-function satisfies the conditions (\ref{ch6}), and,
therefore, gives rise to the CKP tau-function
\beq\label{rat4}
\tau (x, {\bf t}_{\rm o})=\Biggl (\det_{n\times n}
\Bigl (xI-X_0 +\! \sum_{k\geq 1, \,\, k  \, {\rm odd}}
\! kt_k L_0^{k-1}\Bigr )\Biggr )^{1/2},
\eeq
Indeed, we have
$$
\p_{t_{2m}}\log \tau^{\rm KP}\Biggl |_{t_{2k}=0}=
\p_{t_{2m}}\mbox{tr} \log \, \Bigl (xI-X_0 +\sum_{k\geq 1}kt_k L_0^{k-1}\Bigr )
\Biggl |_{t_{\rm e}=0}
$$
$$
=2m \, \mbox{tr}\left [ L_0^{2m-1}\Bigl ((x+t_1)I-X_0 +
3t_3 L_0^2 +5t_5L_0^4+\ldots \Bigr )^{-1}\right ]=0
$$
for all $m\geq 1$ because $\mbox{tr}\, L_0^{2l-1}=0$ for all $l\geq 1$.

\section{Concluding remarks}

The main result of this paper is the identification of the CKP hierarchy as the 
hierarchy of {\it odd times} flows of the KP hierarchy restricted onto the 
locus of its turning points. It suggests that a similar result might be 
valid for the BKP hierarchy. Namely, we conjecture that the 
BKP hierarchy can be identified with the restriction of odd 
times flows of the KP hierarchy onto the locus which in the Sato formulation 
is defined by the equation
\beq\label{conc1}
(\LL^3)_+=\p_x^3+6u\p_x,
\eeq
i.e. the coefficient at the zero power of $\p_x$ in $\LL^3$ vanishes. 
In terms of the tau-function, this condition means that 
\beq\label{conc2}
(\p_{t_2}+\p_{t_1}^2)\log \tau^{\rm KP}\Bigr |_{{\bf t}_{\rm e}}=0
\eeq
for all $t_1, t_3, t_5, \ldots$.
The latter is an analog of equation (\ref{ch6}) defining turning points of the KP hierarchy.

Another interesting problem we plan to consider in the future is the 
Hamiltonian theory of equations of motion for poles
of elliptic solutions to the CKP hierarchy.  
In section 4 we have derived these equation in two ways. 
First, these equations can be obtained by restricting the higher equations of motion of the 
elliptic Calogero-Moser system onto the locus of its 
turning points. As a corollary of this, we have presented  
solutions of these equations in the implicit function form using theta-function 
of the spectral curve. The second approach to the equations of motion is via the 
``generating linear problem'' scheme which allows us to
define the corresponding spectral curve and to prove that it is time-independent
in a direct way (i.e. without any reference to the elliptic Calogero-Moser system).

As it was shown earlier in \cite{KN}, the phase space of the elliptic 
CKP system can be identified with the total space of the Prym varieties 
bundle over the space of the spectral curves. Under this identification the equations of motion become linear on the fibers. Such picture is characteristic for algebraically 
integrable Hamiltonian systems. However, the authors's attempts 
to find the corresponding  Hamiltonian formulation of  equations (\ref{ca1a})  
by a direct guess or by more advanced machinery proposed
in \cite{kp1,kr-nested,kp2} have failed so far.

\section*{Acknowledgments}

\addcontentsline{toc}{section}{\hspace{6mm}Acknowledgments}

We thank S. Natanzon for discussions.
The research has been funded within the framework of the
HSE University Basic Research Program and the Russian Academic Excellence Project '5-100'.

%\section*{Appendix A: How conditions (\ref{ch5}) follow from (\ref{ch6})}
\section*{Appendix A: Proof of Lemma \ref{proposition:even}}
\def\theequation{A\arabic{equation}}
\setcounter{equation}{0}

\addcontentsline{toc}{section}{\hspace{6mm}Appendix A}

%\subsection*{How conditions (\ref{ch5}) follow from (\ref{ch6})}

In this appendix we give a sketch of proof of Lemma \ref{proposition:even},
i.e. we are going to prove that the conditions (\ref{ch5})
and
\beq\label{A01}
\p_x\p_{t_4}\log \tau^{\rm KP}\Bigr |_{{\bf t}_{\rm e}=0}
=\p_x\p_{t_6}\log \tau^{\rm KP}\Bigr |_{{\bf t}_{\rm e}=0}
=\ldots =0.
\eeq
follow from the constraint
\beq\label{A01a}
\p_{t_2}\log\tau^{\rm KP}\Bigr |_{{\bf t}_{\rm e}=0}=0
\quad \mbox{for all
$t_1, t_3, t_5, \ldots $}
\eeq
(see (\ref{ch6}))
provided $\tau^{\rm KP}$ is a KP tau-function, i.e. satisfies all
the equations of the KP hierarchy.

We use the representation of the KP hierarchy in the unfolded form
suggested in \cite{Natanzon1,Natanzon2}, see also section 3.2 of \cite{NZ16}.
Set $F=\log \tau^{\rm KP}$ and $F_{k_1,\ldots ,\, k_m}=\p_{t_{k_1}}\ldots \p_{t_{k_m}}\! F$.
Then the KP hierarchy can be written in the form
\beq\label{A02}
F_{k_1,\ldots , \, k_m}=\sum_{n\geq 1}\sum R_{k_1 ,\ldots ,\, k_m}^{(n)}
\! \left (\begin{array}{lll} s_1 & \ldots & s_n \\
r_1 & \ldots & r_n \end{array} \right )
\p_x^{r_1}F_{s_1}\ldots \p_x^{r_n}F_{s_n},
\eeq
where $m\geq 2$ and $\displaystyle{R_{k_1, \ldots ,\, k_m}^{(n)}
\! \left (\begin{array}{lll} s_1 & \ldots & s_n \\
r_1 & \ldots & r_n \end{array} \right )}$ are universal rational coefficients.
The second sum is taken over all matrices
$\displaystyle{\left (\begin{array}{lll} s_1 & \ldots & s_n \\
r_1 & \ldots & r_n \end{array} \right )}$ such that $s_i, r_i \geq 1$ with the conditions
\beq\label{A03}
\sum_{i=1}^n (s_i+r_i)=\sum_{i=1}^m k_i, \qquad
\sum_{i=1}^n r_i \geq n+m-2.
\eeq
For example \cite{Natanzon1},
\beq\label{A03a}
F_{2,3}=\frac{3}{2}\, \p_xF_4 -\frac{3}{2}\, \p_x^3F_2 -3\p_x F_2 \, \p_x^2F.
\eeq

From the fact that if $\tau^{\rm KP}(x, {\bf t})$ is a tau-function, then
$\tau^{\rm KP}(-x, -{\bf t})$ is a tau-function, too
(this is a corollary of the Hirota equations), it follows that
\beq\label{A04}
\mbox{if $\displaystyle{\sum_{i=1}^n (r_i-1)-m\equiv 1}$ (mod 2), then
$\displaystyle{R_{k_1, \ldots ,\, k_m}^{(n)}
\! \left (\begin{array}{lll} s_1 & \ldots & s_n \\
r_1 & \ldots & r_n \end{array} \right )=0.}$}
\eeq

First we prove (\ref{A01}). The proof is by induction.
We assume that (\ref{A01}) is true
for $\p_xF_{2}, \ldots , \p_xF_{2k}$ (this is certainly true if $k=1$) and will
deduce from (\ref{A02}) that it is true for $k\to k+1$.
From (\ref{A01a}) and (\ref{A02}) at $m=2$ we have:
\beq\label{A02a}
\begin{array}{c}
\displaystyle{
0=F_{2, \, 2k+1}=\sum_{s_1+r_1=2k+3}R^{(1)}_{2, \, 2k+1}
\left (\begin{array}{c} s_1\\r_1\end{array}
\right ) \p_x^{r_1}F_{s_1}}
\\ \\
\displaystyle{+
\sum_{s_1+s_2+r_1+r_2=2k+3}R^{(2)}_{2, \, 2k+1}
\left (\begin{array}{cc} s_1&s_2\\r_1&r_2\end{array}
\right ) \p_x^{r_1}F_{s_1}\p_x^{r_2}F_{s_2}+\ldots}
\end{array}
\eeq
Separating the term with $r_1=1$ in the first sum in the right hand side of
(\ref{A02a}), we write it as
\beq\label{A02b}
0=F_{2, \, 2k+1}=R^{(1)}_{2, \, 2k+1}
\left (\begin{array}{c} 2k+2\\1\end{array}
\right ) \p_x F_{2k+2}  \,\, +\;
\mbox{all the rest}.
\eeq
Now, recalling the condition (\ref{A04}), we see that the non-zero coefficients at the
different terms in the right hand side are when
$\displaystyle{\sum_{i=1}^n s_i =n-1 \; \mbox{(mod $2$)}}$. From this it follows that
for both odd and even $n$ at least one of the $s_i$'s must be even (and less then
$2k+2$). Therefore, ``all the rest'' terms vanish by the induction assumption.
Since the coefficient $\displaystyle{R^{(1)}_{2, \, 2k+1}
\left (\begin{array}{c} 2k+2\\1\end{array}
\right )}$ is not equal to zero (see \cite{Natanzon1}), we conclude from (\ref{A02b})
that $\p_x F_{2k+2}=0$.

Next we are going to prove that if $\p_xF_{2k}=0$
for all $k\geq 1$ and all $t_1, t_3, \ldots$, then
$F_{k_1,\ldots ,\, k_m}=0$ for all even $k_1, \ldots , k_m$ and odd $m\geq 3$. As soon as $m+1$
and all $k_i$'s are even, we can, using (\ref{A03}), rewrite the condition (\ref{A04})
in the form
\beq\label{A05}
\sum_{i=1}^n s_i \equiv n \,\,\, \mbox{(mod 2)}.
\eeq
But if at least
one of $s_i$ in (\ref{A02}) is even, then the corresponding term vanishes because
$F_{2k}=0$ for all $k\geq 1$. Therefore, all the $s_i$'s must be odd, i.e.,
$s_i=2l_i+1$ and so the condition (\ref{A05}) is satisfied which means that the
coefficient $\displaystyle{R_{k_1, \ldots ,\, k_m}^{(n)}
\! \left (\begin{array}{lll} s_1 & \ldots & s_n \\
r_1 & \ldots & r_n \end{array} \right )}$ vanishes. This proves that $F_{k_1, \ldots ,\, k_m}=0$.

\section*{Appendix B: Proof of equation (\ref{ca5})}
\def\theequation{B\arabic{equation}}
\setcounter{equation}{0}

\addcontentsline{toc}{section}{\hspace{6mm}Appendix B}

%\subsection*{Proof of equation (\ref{ca5})}

Here we prove the matrix identity (\ref{ca5}).

First of all we note that $\dot L_{ik}=-(\dot x_i-\dot x_k)\Phi '(x_i-x_k)$, and, therefore,
we have $\dot L =-[\dot X, B]$. To transform the commutators
$[L,B]+[L,D]$, we use the identity
\beq\label{A1}
\Phi (x )\Phi '(y)-\Phi (y)\Phi '(x)=\Phi (x+y)(\wp (x) -\wp (y)).
\eeq
With the help of it we get for $i\neq k$
$$
-\Bigl ([L,B]+[L,D]\Bigr )_{ik}
$$
$$
=\, \sum_{j\neq i,k}\Phi (x_i-x_j)\Phi '(x_j-x_k)-
\sum_{j\neq i,k}\Phi ' (x_i-x_j)\Phi (x_j-x_k)
$$
$$
+\, \Phi (x_i-x_k)\Bigl (\sum_{j\neq k}\wp (x_j-x_k)-\sum_{j\neq i}\wp (x_i-x_j)\Bigr )=0,
$$
so we see that
$[L,B]+[L,D]$ is a diagonal matrix. To find its matrix elements, we use the limit
of (\ref{A1}) at $y=-x$:
$$
\Phi (x)\Phi '(-x)-\Phi (-x)\Phi '(x)=\wp '(x)
$$
which leads to
$$
-\Bigl ([L,B]+[L,D]\Bigr )_{ii}
$$
$$
=\,
\sum_{j\neq i}\Bigl (\Phi (x_i-x_j)\Phi ' (x_j-x_i)-\Phi ' (x_i-x_j)\Phi (x_j-x_i)\Bigr )
=\sum_{j\neq i}\wp '(x_i-x_j)=D'_{ii},
$$
so we finally obtain the matrix identity
\beq\label{A4}
[L,B]+[L,D]=-D'.
\eeq

Combining the derivatives of (\ref{A1}) w.r.t. $x$ and $y$, we obtain the
identity
\beq\label{A5}
\Phi (x)\Phi ''(y)-\Phi (y)\Phi ''(x)=2\Phi '(x+y)(\wp (x)-\wp (y))
+\Phi (x+y)(\wp '(x)-\wp '(y))
\eeq
which allows us to prove the matrix identity
\beq\label{A9}
[L,C]=-2[D, B]+D'L+LD',
\eeq
which
is used, together with (\ref{A4}), to transform $\dot L+[L,M]$ to the form (\ref{ca5}).

\end{document}